\documentclass[graybox]{svmult}

\usepackage{helvet}         % selects Helvetica as sans-serif font
\usepackage{courier}        % selects Courier as typewriter font
\usepackage{type1cm}        % activate if the above 3 fonts are
\usepackage{makeidx}         % allows index generation
\usepackage{graphicx}        % standard LaTeX graphics tool
\usepackage{multicol}        % used for the two-column index
\usepackage[bottom]{footmisc}% places footnotes at page bottom
\usepackage{amsmath}

\newcommand{\be}{\begin{eqnarray}}
\newcommand{\ee}{\end{eqnarray}}
\newcommand{\non}{\nonumber\\}

\newcommand{\ave}[1]{\left\langle #1 \right\rangle}

\newcommand{\rp}{\Psi_{RP}}

\makeindex             % used for the subject index
\begin{document}

\title*{Charge-Dependent Correlations in Relativistic 
Heavy Ion Collisions and the Chiral Magnetic Effect}
\titlerunning{Charge-Dependent Correlations}
\author{Adam Bzdak, Volker Koch, Jinfeng Liao}
\institute{Adam Bzdak \at RIKEN BNL Research Center, Brookhaven National Laboratory,   Upton, NY 11973, USA.\\ \email{abzdak@bnl.gov}
\and Volker Koch \at Nuclear Science Division, Lawrence Berkeley
National Laboratory, MS70R0319, 1 Cyclotron Road, Berkeley,
California 94720, USA.\\ \email{vkoch@lbl.gov}
\and Jinfeng Liao \at Physics Department and Center for Exploration of Energy and Matter, Indiana University, 2401 N Milo B. Sampson Lane, Bloomington, IN 47408, USA.\\
RIKEN BNL Research Center, Brookhaven National Laboratory,
   Upton, NY 11973, USA.  \\  
   \email{liaoji@indiana.edu}
}
\maketitle

\abstract*{We provide a phenomenological analysis of present experimental searches for local parity violation
  manifested through the Chiral Magnetic Effect. We introduce and  discuss the  relevant
  correlation functions used for the measurements. Our 
  analysis of the  available data from both RHIC and LHC shows that
  the present experimental evidence for the Chiral Magnetic Effect is rather
  ambiguous. We further discuss in some detail various background
  contributions due to conventional physics, which need to be
  understood quantitatively in order to draw a definitive conclusion
  about the existence of local parity violation in heavy ion collisions.}

\abstract{We provide a phenomenological analysis of present experimental searches for local parity violation
  manifested through the Chiral Magnetic Effect. We introduce and  discuss the  relevant
  correlation functions used for the measurements. Our 
  analysis of the  available data from both RHIC and LHC shows that
  the present experimental evidence for the Chiral Magnetic Effect is rather
  ambiguous. We further discuss in some detail various background
  contributions due to conventional physics, which need to be
  understood quantitatively in order to draw a definitive conclusion
  about the existence of local parity violation in heavy ion collisions.}

\vspace{2.4cm}
{\small{
\noindent To appear in Lect. Notes Phys. "Strongly interacting matter in magnetic
fields" (Springer), edited by D. Kharzeev, K. Landsteiner, A. Schmitt,
H.-U. Yee }}

\newpage

\section{Introduction}

The theoretical study of topological solitons in field theories has a long history. 
Quite generically, these objects arise as solutions to the classical equations of motion 
for field theories due to the nonlinearity of the equations as well as
due to specific boundary conditions. 
They are found in field theories of various dimensions (2D kinks, 3D monopoles, 4D instantons), 
and are known to be particularly important in the non-perturbative domain where the theories are strongly coupled. For a recent review, see e.g. \cite{'tHooft:1999au}.

Topological objects in Quantum Chromodynamics (QCD) are known to play important roles 
in many fundamental aspects of QCD \cite{'tHooft:1999au}. For example, instantons are 
responsible  for various properties of the QCD vacuum, such as
spontaneous breaking of chiral symmetry and the $U_A(1)$ anomaly
(see e.g. \cite{Schafer:1996wv,ES_book}). Magnetic
monopoles, on the other hand, are speculated to be present in the
QCD vacuum in a Bose-condensed form which then enforce the color
confinement, known as the dual superconductor model for QCD
confinement, which is strongly supported by lattice
QCD calculations (see e.g. \cite{Ripka:2003vv,Bali:1998de}).
Alternatively vortices are believed to describe the
chromo-electric flux configuration (i.e. flux tube) between a
quark-anti-quark pair in the QCD vacuum which in turn gives rise
to the confining linear potential (see e.g. reviews in
\cite{Bali:1998de,Greensite:2003bk}). Some of these objects,
such as monopoles \cite{Liao:2006ry} and flux tubes
\cite{Liao:2007mj}, may also be important degrees of freedom in
the hot and deconfined QCD matter close to the transition
temperature $T_c$, and may be responsible for the observed
properties of the so called strongly coupled quark-gluon plasma
\cite{Shuryak:2008eq,Liao:2008dk}.

Since the existence of such topological objects is theoretically well
motivated and their effects on the dynamics are deemed to be important, 
a direct experimental detection of such 
objects or at least of certain unique imprints by them, would be a highly 
desirable goal.
This review will discuss recent efforts and progress toward that
goal, specifically in the context of relativistic heavy ion collisions 
through the measurement and analysis of charge-dependent 
correlations.

\subsection{The Chiral Magnetic Effect in brief}

An interesting suggestion by Kharzeev and
collaborators
\cite{Kharzeev:2004ey,Kharzeev:2007tn,Kharzeev:2007jp,Fukushima,Buividovich:2009wi,Kharzeev:2009fn} on the direct manifestation of effects from topological objects is
the possible occurrence  of $\cal P$- and $\cal CP$-odd (local)
domains due to the so-called sphaleron or anti-sphaleron  transitions in the hot
dense QCD matter created in relativistic heavy ion collisions. 
Imagine that in a single event created in a heavy
ion collision the gauge field configurations in the space-time zone of the created hot dense matter experience a single sphaleron transition. As a result this 
local zone acquires a non-zero topological charge which is parity-odd. This non-zero topological 
charge, when coupled with light quarks through the triangle anomaly,
induces a non-zero chirality  for the
quarks. In other words it generates an imbalance between left- and
right-handed quark numbers, or a non-zero axial charge density. To be
precise, there is no violation of parity at the interaction level, but
rather a local creation of matter with non-zero axial charge
density, which is a $\cal P$- and $\cal CP$-odd quantity. 

A concrete proposal for experimental detection is the so-called Chiral
Magnetic Effect (CME) \cite{Kharzeev:2007jp}. The effect itself states
that in the presence of external electromagnetic (EM) magnetic field
$\vec B$, a nonzero axial charge density will lead to an EM electric current
along the direction of the magnetic field $\vec B$:
\begin{eqnarray}  \label{cme}
\vec j_V={N_c\ e \over 2\pi^2} \mu_A \vec B    
\end{eqnarray} 
where $\mu_A$ is the axial chemical potential associated with the non-zero axial charge
density present in the system, and $N_{c}$ is the number of colors. This elegant relation is theoretically
well established  in both the weakly-coupled and the
strongly-coupled regimes of the theory as will be discussed in several contributions to this volume.

At first sight, it might seem that the above relation is violating
parity: under spatial rotation and inversion the EM electric current
$\vec j_V$ transforms like a vector,  while the magnetic field, $\vec
B$, transforms like an axial- or pseudo-vector. Therefore, the factor in
Eq. (\ref{cme}) relating the two will have to be parity-odd. This is
indeed the case, since  $\mu_A$ that enters the above relation is a
pseudo-scalar quantity which changes sign under parity
transformation. Thus the CME relation, Eq. (\ref{cme}), is invariant
under parity transformation. However, in a region with 
nonzero, either positive or
negative,  $\mu_A$ certain parity-odd observables, e.g. the 
pesudoscalar quantity $<\vec j_V \cdot \vec B>$, may acquire nonzero expectation values.
It is only in this sense that one may refer to it as ``local parity violation''.

In addition there is a complimentary relation, as one might have  guessed
from the ``duality'' by interchanging the roles of V (vector) and A (axial), that has
been called the Chiral Separation Effect (CSE). The CSE  refers to the
separation of chiral (or axial) charge along the axis of the external
EM magnetic field at finite density of the vector charge, for example
at finite baryon number density \cite{son:2004tq,Metlitski:2005pr}. The resulting axial current is given by
\begin{eqnarray} \label{cse}
\vec j_A={N_c\ e\over 2\pi^2} \mu_V \vec B 
\end{eqnarray}
with the $\mu_V$ here being the  baryon number chemical
potential.  Furthermore the combination of the two effects, CME and CSE, 
gives rise to an interesting propagating collective mode: the vector
density induces an axial current which transports and creates a locally
nonzero axial charge density, which in turn leads to a vector current
that further transports and creates a locally nonzero vector density,
and so on. This is called Chiral Magnetic Wave (CMW) \cite{Kharzeev:2010gd}, just like
Maxwell's electromagnetic waves represent the coupled evolution of the
electric and magnetic fields. The CMW is a general
concept that includes both the CME and CSE
effects. It is robust in the sense that it takes the form of a collective excitation like the sound wave without relying on a quasi-particle picture. 

We end the general introduction with two comments: first, the CME in
the language of CMW  induces a charged dipole (of the
vector density distribution) that results from an initial nonzero
axial charge density; second it has been recently pointed
\cite{Burnier:2011bf} that an initial vector charge density via CMW
will lead to a charged quadrupole distribution that may be observable
in heavy ion collisions. 
For the rest of this contribution we will focus on the charged dipole signal for the CME phenomenon. 

\subsection{Hunting for the CME in heavy ion collisions}

Now we turn to two key questions: can the Chiral Magnetic Effect occur
in heavy ion collisions, and if so, what observables serve as
unambiguous signals for the CME? 

The answer to the first question seems to be positive. Two elements are needed
for the CME to occur: an external magnetic field and a locally nonzero
axial charge density. The relativistically moving heavy ions,
typically with large positive charges (e.g. $+79e$ for Au), carry
strong magnetic (and electric) fields with them. In the short moments
before/during/after the impact of two ions in non-central
collisions, there is a very strong magnetic field in the reaction
zone \cite{Rafelski:1975rf,Kharzeev:2007jp}. In fact, such a magnetic field is 
estimated to be of the order of $m_{\pi }^{2}\approx 10^{18}$ Gauss
\cite{Skokov:2009qp,Bzdak:2011yy} (see also \cite{Deng:2012pc}), probably the strongest, albeit transient, 
magnetic field in the present Universe. 
The other required element, a locally non-vanishing axial charge
density, can also be created in the reaction zone during the collision
process through sphaleron transitions (see e.g. \cite{Kharzeev:2009fn} for disucssions and references therein). 
As such, it appears at least during the very early stage of a heavy
ion collision, there can be both strong magnetic field and nonzero
axial charge density in the created hot matter. Therefore, the CME
should take place, that is, an electric current will be generated
either parallel or anti-parallel to the magnetic field $\vec{B}$
depending on the axial charge density is positive (due to sphaleron)
or negative (due to anti-sphaleron). How large this current is, is of
course another question, see e.g. \cite{Muller:2010jd}.

The answer to the second question is much more difficult.  Extracting  the effects of the CME, which most
likely occur at the very early stage of the collision, from the final
observed hadrons, involves  many uncertainties. 
First, it is quite unclear how long the magnetic field could remain
strong: while the peak value is large, it decays very rapidly with
time (if the only source of such field is from the protons in the
ions) \cite{Toneev:2011aa}. Second, if the CME current is generated
mostly at very early time, it is not clear to which extent this
current could survive  without significant modifications, since we
know that the created quark-gluon plasma behaves like a strongly
interacting fluid. Furthermore, even if this current survives, one has
to find the right observable for its detection. 
At present, there is no satisfactory resolution on the first two
issues. This will likely require comprehensive and quantitative model studies. In this review we will only focus on the third issue --- the observables to be used for measuring the possible CME current and related ``background'' effects.

In a simplistic view, one may
  consider the ultimate manifestation of the CME as a separation of
  charged hadrons along the direction of the initial magnetic field:
  more positive hadrons moving in one  direction while more negative
  hadrons in the opposite direction.  As a result, the momentum
  distribution of the final hadrons will have a charged dipole
  moment. 
The direction of such a momentum space dipole is expected to
be along the $\vec B$ field, parallel or anti-parallel, depending 
on the sign of the initial axial charge density in a given
event. Since the initial axial charge may be positive and negative
with equal probability, the event average of the momentum space
dipole vanishes,  $<\vec j_V \cdot \vec B>=0$. This reflects the
fact that parity is not broken globally by the strong interaction, 
so that any pseudo-scalar quantity, such as $<\vec j_V \cdot
\vec B>$,  will have to vanish. What one can hope for, however, is to
measure the fluctuation or variance of this charge separation, i.e.
$<(\vec j_V \cdot \vec B)^2>$, which is a parity even quantity.  As we
will discuss later, the prize one has to pay is that other, conventional
correlations, not related to the CME, may contribute to observables
which are sensitive to the variance of charged dipole moment. 

Recently the STAR collaboration at 
the Brookhaven's Relativistic Heavy Ion
Collider (RHIC) has reported \cite{Star:2009uh} first measurements of a charge dependent correlation function in heavy ion collisions, which may by sensitive 
to the Chiral Magnetic Effect. The essential idea of the measurement, 
proposed by Voloshin \cite{Voloshin:2004vk}, is based on two important 
features: first, in non-central heavy ion collisions, the direction of initial strong magnetic field 
is strongly correlated with the so-called reaction plane, which is  spanned by the impact parameter and the beam direction. The $\vec B$ field 
is pointing (mostly) along the normal of reaction plane, albeit with random up/down orientation; 
second, the CME-induced current, or 
the charged dipole in momentum space,  implies particular
charge-dependent correlation patterns. The same-sign charged hadrons
will prefer moving together while the opposite-sign charged hadrons moving back-to-back 
along the $\vec B$ field direction, and thus perpendicular to the
reaction plane, which is commonly referred to as the out-of-plane direction 
\footnote{As a note of caution, the strong correlation between the $\vec
  B$ field direction and the participant-plane are
  considerably  modified when the strong fluctuations in the initial
  condition are properly taken into account. As a result the two are
  rather  weakly  correlated in very central and very peripheral collisions \cite{Bzdak:2011yy,Deng:2012pc,B_fluctuation}. }. 
While these measurements  and their implications will be discussed in
detail in Section~\ref{sec:3}, let us briefly
  summarize the present status: the STAR (later PHENIX, and also
  ALICE) data show very interesting charge dependent azimuthal
  correlation patterns, and some features are in line with the CME
  predictions. Other aspects of the data, on the other hand, are very
  hard to  understand within the framework of the  CME.  At present,
  therefore, the observation of the Chiral Magnetic Effect in
  heavy ion collisions, and the local parity violation in the
  aforementioned sense,  has not been established experimentally, and
  additional measurements as well as further theoretical analysis are
  required before definitive conclusions can be drawn.

\vspace{0.5cm}
This review is organized as follows: in Section 2, we will present
a general discussion on the charge-dependent correlation measurements 
in heavy ion collisions, with the  emphasis on the CME related observables; 
in Section 3, the presently available data from heavy ion collisions at a variety 
of collision energies will be examined and their interpretations will be 
critically evaluated; in Section 4, various possible ``background'' effects  
and their manifestation in various observables will be quantitatively analyzed;  
finally in Section 5 we summarize and conclude.

\section{The charge-dependent correlation measurements}
\label{sec:2}

In this Section, we focus on various charge-dependent correlation 
measurements  in heavy ion collisions and what can be learned 
from these observables. The emphasis will not be on the data
themselves, which will be the subject of the next Section. Instead we will
set up the conceptual framework for studying the azimuthal correlations, 
discuss possible complications in the design of the observables, and examine 
the connection between physical effects and the measurements.  

\subsection{General considerations concerning azimuthal correlation measurements}
\label{sec:2.1}

The basic experimental information about the  (hadronic) final state
of a heavy ion collision 
consists of the momenta and the identity -- the electric
charge, mass and possibly other quantum numbers -- of all hadrons observed in
the acceptance of a given experiment. Customarily, the three-momentum 
$\vec p$ is
represented by the (longitudinal) rapidity, 
$y$, the transverse momentum $p_t$ as
well as the  the azimuthal angle $\phi$. Events may further be grouped
according to the charged particle multiplicity, which is a good
measure of the centrality or impact parameter of a collision. From a
given sample of events one can then extract the single particle distributions,  $d^3N/dydp_t^2 d\phi$ either for 
all charged hadrons or, more selectively,  for identified pions,
kaons, protons, etc. In order to study possible correlations one
analyses two-particle, three-particle and multi-particle distributions
of various kinds.  
Most of the discussion in this review will focus on the
dependence of various measurements on the azimuthal angle. The rapidity $y$
and the transverse momentum $p_t$ will either be in specific bins or integrated over.

The analysis of azimuthal distributions has to deal with the fact that
the azimuthal direction of each collision, characterized by either the
direction of the angular momentum or the impact parameter, is randomly
distributed in the laboratory frame. Therefore, a single particle
azimuthal distribution, $dN/d\phi$ will always be uniform and, thus,
rather meaningless. To learn something about azimuthal distributions,
one either measures distributions of the difference of the azimuthal
angles of two particles, $dN/d\left(\phi_1-\phi_2 \right)$, or one
determines the azimuthal orientation of a given event and studies distributions with
respect to this direction. Commonly the azimuthal
direction of the so-called reaction plane is used to characterize the
orientation of an event. As already discussed in the
Introduction, the reaction plane is spanned by the beam direction and
the impact parameter of the collision. Its orientation in the
laboratory frame is given by the so-called reaction plane angle,
$\Psi_{RP}$, which measures the direction of the impact parameter in
the laboratory frame. Given the reaction plane angle, one then can
study azimuthal angular distributions with respect to the reaction
plane angle, $ f\left(\phi - \Psi_{RP}\right) = dN/d\left(\phi-\Psi_{RP}\right)$. 
Clearly the determination of the reaction plane requires the
measurement of other particles in addition to that used for the angular
distribution (for a comprehensive review, see
\cite{Voloshin:2008dg}). 
Therefore, the extraction of azimuthal distributions will
require the measurement of two-particle (for the angular difference
distribution $dN/d\left(\phi_1-\phi_2 \right)$) or even higher particle distributions. 

\begin{figure}[b]
\begin{center}
\includegraphics[width=8cm]{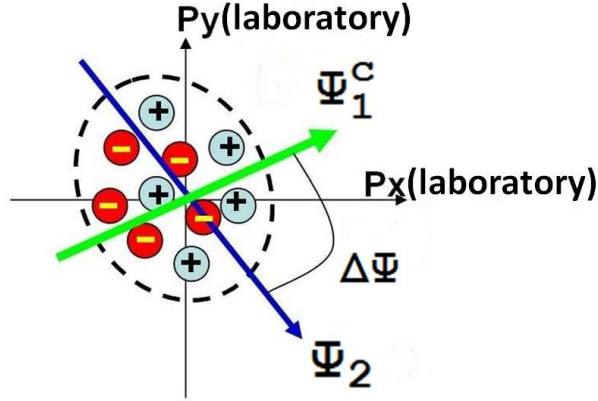}
\end{center}
\caption{A schematic demonstration of the proposed simultaneous
analysis of $\hat{Q}^c_1$ and $\hat{Q}_2$ vectors in the same
event. } \label{fig_demo}
\end{figure}

However, it is important to distinguish between the need to measure
two- or many-particle distributions to study azimuthal distributions,
and the presence of true dynamical two- or many- particle correlations. To make this
distinction more transparent, it is useful to introduce an {\em
  intrinsic} frame or coordinate
system where the $x$-direction is given by the direction of the impact
parameter, which is typically referred to as the so-called
``in-plane-direction'', and the $y$ direction is defined by the
angular momentum, or the so-called ``out-of plane direction''. The
relative angle of the x-axis of the intrinsic frame and that of the laboratory
frame is then given by the reaction plane angle $\Psi_{RP}$, as illustrated in Fig.\ref{fig_demo}. In theoretical
considerations and model calculations  the orientation of the reaction
plane is  assumed to be known, or in other words, these calculations
take place in the intrinsic frame. Finally, the azimuthal
angle $\Phi$ in the intrinsic frame is related to the laboratory angle $\phi$
by
\be 
\Phi=\phi-\Psi_{RP}
\ee

To continue, 
let us, as an example, consider a single particle distribution in the intrinsic frame
\begin{eqnarray} \label{eq_f_v2}
f_1(\Phi)=f_1(\phi-\Psi_{RP}) \propto 1 + 2 v_2 \cos[2(\Phi)]=1 + 2 v_2 \cos[2(\phi-\Psi_{RP})]
\label{single-particle-lab}
\end{eqnarray}
which has an azimuthal asymmetry, characterized by the second Fourier
component of strength $v_2$. This kind of distribution, which will be
relevant for the subsequent discussion, is important in
the context of the observed azimuthal asymmetries in heavy ion
collisions, which are generally attributed to the hydrodynamics
evolution of the system in non-central collisions. The parameter $v_2$
is commonly referred to as the elliptic flow coefficient. For a
detailed discussion see \cite{Voloshin:2008dg}.  
The value for the elliptic flow parameter, $v_2$, may be obtained by
 measuring the second moment of the angular distribution, $\langle \cos2\left(\phi-\Psi_{RP}\right)\rangle$ . To this
end we have to determine the reaction plane angle in each event, 
calculate the average moment in the intrinsic frame of each event and then average over events:
\begin{eqnarray}
&& \langle \cos\left[2 \left(\phi-\Psi_{RP}\right)\right]\rangle  \nonumber \\
&& = \frac{1}{N_{\rm events}}\sum_{{\rm event} \,i=1}^{N_{\rm events}}  {\bigg \{}
\frac{1}{N(i)}\sum_{{\rm particle}\, k=1}^{N(i)}
\cos\left[2\left(\phi_k -\Psi_{RP}(i)\right)\right] {\bigg \}}
\label{eq:v2_real}
\end{eqnarray}
In terms of the distribution function $f_1$ this can be expressed
as\footnote{In reality the ability to express the actual measurement,
  as described in Eq.~(\ref{eq:v2_real}), in terms of an average of 
  moments of the intrinsic distribution over the reaction plane angle
  requires a detailed analysis of all non-flow effects and flow
  fluctuations, as discussed in detail in Ref. \cite{Voloshin:2008dg}.}  
\be
\langle \cos\left[2 \left(\phi-\Psi_{RP}\right)\right]\rangle = 
\frac{\int d\Psi_{RP}  \int d\phi \,f_1(\phi-\rp)
  \cos\left[2 \left(\phi-\rp\right)\right]}{\int d\Psi_{RP}  \int
  d\phi  \,f_1(\phi-\rp)}
\label{eq:v2_simple}
\ee

Let us next consider the two-particle distribution 
\be 
f_2(\Phi_1,\Phi_2) = f_1(\Phi_1) f_1(\Phi_2) +
C\left(\Phi_1,\Phi_2\right)
\label{eq_f_12}
\ee
where the first term is simply the product of the single particle
distributions, and the second term,   $C\left(\Phi_1,\Phi_2\right)$
represents possible, {\em true}, two-particle  correlations. 
Since the two-particle distribution depends on two angles, $\Phi_1$ and
$\Phi_2$, in general it will have terms which depend only on the difference
of the angle $\sim (\Phi_1-\Phi_2)=(\phi_1-\phi_2)$, and which are
independent of the direction of the reaction plane. It will also have
terms which depend on the sum of the angles, 
$\sim (\Phi_1+\Phi_2)=(\phi_1+\phi_2-2\rp)$ which are dependent on the
reaction plane direction. This may be illustrated by inserting into 
Eq. (\ref{eq_f_12}) the single particle distribution, Eq. (\ref{eq_f_v2}), and
neglecting the correlation term, i.e., setting  $C(\Phi_1,\Phi_2)=0$. In this case
\be
f_2(\Phi_1,\Phi_2) &=& f_1(\Phi_1) f_1(\Phi_2)\non
&\sim& 2v_2^2 \cos[2(\Phi_1-\Phi_2)] + 2v_2^2 \cos[2(\Phi_1+\Phi_2)]\non
&=& 2v_2^2 \cos[2(\phi_1-\phi_2)] + 2v_2^2
\cos[2(\phi_1+\phi_2-2\Psi_{RP})]
\label{eq:f2_example}
\ee
The term  $\sim  \cos[2(\phi_1-\phi_2)]$ which depends on the
difference of the angles can then be extracted by the measurement of
the two-particle correlation  
\begin{eqnarray} \label{eq_cos_2phi}
\ave{\cos[2(\phi_1-\phi_2)]} \sim \int_{\Phi_1} \int_{\Phi_2}
f_2(\Phi_1,\Phi_2) \cos[2(\phi_1-\phi_2)]
\end{eqnarray} 
The measurement of the term $\sim \cos[2(\phi_1+\phi_2-2\Psi_{RP})]$ requires the
determination of the reaction plane, or at least a three-particle
correlation measurement.
For our example, Eq. (\ref{eq:f2_example}),
$\ave{\cos[2(\phi_1-\phi_2)]} \sim v_2^2$, and in fact this is one of
the frequently used (and the simplest) methods for measuring the
elliptic flow. However, this method suffers from the so-called ``non-flow''
\cite{Voloshin:2008dg} contributions, which are due to the correlation term
we have neglected in our example. 
Our simple example also demonstrates a very important fact: single
particle distributions, 
such as $f_1$ do contribute to multi-particle azimuthal correlations. This
will be essential for the subsequent discussion where one of the tasks will be
to disentangle the effects from true correlations and contributions from the
single particle distributions.

The above discussion can be easily extended to three- (and more) particle
densities with the same basic conclusions: 
\begin{itemize}
\item The n-particle density
will have terms which do not depend on the reaction plane, and thus
may be extracted by the measurement of appropriate n-particle
correlations. It will also have reaction plane dependent terms, which
require the measurement of at least n+1 particle correlations or the
determination of the reaction plane. 
\item 
Unless not very carefully designed, multi-particle correlations will
contain contributions from the single particle distribution.
\end{itemize}
Finally, the measurement of angular correlations is of course not
restricted to the second Fourier moment. Recently the harmonic
moments, $\ave{\cos[n(\phi_1-\phi_2)]}$, have been measured
 in order to study flow fluctuations \cite{Harmonics}. These
 correlations may also be measured in a more selective way, such as
 correlations for particles with same or opposite electric charges
 (the charge-dependent correlations), correlations for particles with
 certain quantum numbers (e.g. baryon-strangeness \cite{Koch:2005vg}), or correlations
 for particles within or between certain kinematic regions (e.g. the soft-hard 
 correlations \cite{Zhang:2012mi}), etc.

\subsection{Measuring the charge separation through azimuthal correlations}
\label{sec:2.2}

Let us turn to possible azimuthal correlation measurements as the
signal for the Chiral Magnetic Effect. Specifically, as discussed in
the Introduction, we have to find azimuthal correlations which are
sensitive to a possible out-of-plane ``charge separation''.

We begin by defining what we mean by ``charge separation
effect''. Consider the distribution of final state
hadrons in the transverse momentum space as schematically shown in
the Fig. \ref{fig_demo}. If the ``center'' of the positive charges
happens to be different from that of the negative charges, then
there is a separation between two types of charges which may be
quantified by an ``electric dipole moment'' in the transverse
momentum space. Such a separation may arise either simply from
statistical fluctuations or may be due to specific dynamical effect such
as the CME. We note that such a charge separation occurs already  at
the single-particle distribution level in the  {\em
intrinsic} frame. Let us, therefore, define a 
 charge-dependent single-particle azimuthal distribution, which, besides a possible
momentum-space electric dipole moment, also includes the
presence of elliptic flow: 
\begin{eqnarray}\label{eqn_dist}
f_\chi\,(\phi,q) \propto 1+2\, v_2\cos[2(\phi-\rp)]+2\, q
\, \chi\, d_1\cos(\phi-\Psi_{CS})
\end{eqnarray}
Here $q$ and $\phi$ represent the charge and the azimuthal angle
of a particle, respectively. The parameters $v_2$ and $d_1$ quantify
the elliptic flow and the charge separation effect, while
$\Psi_{CS}$ specifies the azimuthal orientation of the
electric-dipole 
and $\Psi_{RP}$ the direction of the reaction plane (see Fig. \ref{fig_demo}). 
It is important to
notice that an additional random variable $\chi=\pm 1$ is
introduced. This  accounts for the fact that in a given event we
may have sphaleron or anti-sphaleron transitions resulting in
charge separation parallel or anti-parallel to the magnetic field.
Consequently the sampling over all events with a given reaction
plane angle, $\Psi_{RP}$, corresponds to averaging  the
intrinsic distribution $f_\chi$ over $\chi$, namely
$f_{1}=<f_\chi>_\chi \propto 1+2\, v_2\cos(2\phi-2\Psi_{RP})$.
Physically speaking this means that the charge separation (or electric
dipole, being $\cal P$-odd) flips sign randomly and averages to zero, thus
causing the expectation value of any parity-odd operator to
vanish. However, since $<{\chi^2}> =1$ the presence
 of an event-by-event electric dipole may be observable through its
 variance. 

For measurements related to heavy ion collisions one may
reasonably assume particle charges to be $|q|=1$ which is the case
for almost all charged particles, e.g., charged pions and kaons,
protons, etc. We note, that the  above distribution does
{\em not} contain a directed flow term $\sim \cos(\phi-\rp)$
for either type of charges, which 
is reasonable if the distribution is measured in a symmetric rapidity
bin. 

Finally one may also consider a $p_t$-differential formulation of
the charge separation effect or charge separation effects
associated with higher harmonics in the azimuthal angle $\phi$. We note that the charge separation term considered in Eq. (\ref{eqn_dist}) is actually the
lowest harmonic in a more general charge-dependent Fourier series expansion
in terms of the azimuthal angle. Various higher harmonics may  
be present due to e.g. the occurrence of multiple topological objects and their
distributions over the entire transverse plane, the influence of
transverse flow as well as the re-scattering of the CME current with medium.
Here we concentrate the discussion
on a possible measurement of the lowest harmonic that is most relevant to 
the CME current.

Let us next discuss how the above defined  charge-dependent
intrinsic single-particle distribution contributes to the charge-dependent azimuthal correlations recently measured by the STAR collaboration
in \cite{Star:2009uh}. Note that here we are only considering the contribution 
from the charge separation term in Eq. (\ref{eqn_dist}), while there are certainly additional contributions from two- and multi-particle
correlations which we will discuss later in Section 4. Specifically the STAR
collaboration has measured the following two- and
three-particle correlations \cite{Star:2009uh}. 

(i) The two-particle correlation $<\cos(\phi_i-\phi_j)>$ for
same-charge pairs ($++/--$) and opposite-charge pairs ($+-$). The
contribution to this correlator due to the charge-dependent 
intrinsic single-particle distribution, Eq. (\ref{eqn_dist}) is:
\begin{eqnarray}
&&\label{eqn_s_ij}  \delta_{++/--}\equiv <\cos(\phi_i-\phi_j)>_{++/--}\, =d_1^2   \\
&&\label{eqn_o_ij}  \delta_{+-} \equiv <\cos(\phi_i-\phi_j)>_{+-}\, =-d_1^2   
\end{eqnarray}

(ii) The three-particle correlation
\mbox{$<\cos(\phi_i+\phi_j-2\phi_k)>$} for same-charge pairs
($i,j=++/--$) and opposite-charge pairs ($i,j=+-$) with the third
particle, denoted by index $k$, having any charge. The
contribution to these correlators due to the distribution,
Eq. (\ref{eqn_dist}), turns out to be
\begin{eqnarray}
&& <\cos(\phi_i+\phi_j-2\phi_k)>_{++/--,\, k-any}\, = v_2\, d_1^2
\cos(2\, \Delta\Psi_{CS}) \\
&& <\cos(\phi_i+\phi_j-2\phi_k)>_{+-,\, k-any}\, = -v_2\, d_1^2
\cos(2\, \Delta\Psi_{CS})
\end{eqnarray}
where  ``k-any'' indicates that  the charge of the 3-rd particle
may assume any value/sign. We have also defined the relative angle of the
charged dipole with respect to the reaction plane, $\Delta\Psi_{CS} \equiv
\Psi_{CS}-\Psi_{RP}$. The purpose of correlating the charged pair with 
the third particle is to address the reaction plane
dependence of the pair-distribution, as discussed in the previous
Section, Sect. \ref{sec:2.1}. Indeed, the STAR collaboration has demonstrated
\cite{Star:2009uh} that the above three particle correlator is
dominated by the reaction plane dependent two-particle
correlation function $<\cos(\phi_i+\phi_j-2\Psi_{RP})>$
and within errors they have found that
\begin{eqnarray}
<\cos(\phi_i+\phi_j-2\phi_k)> = v_2
<\cos(\phi_i+\phi_j-2\Psi_{RP})>
\end{eqnarray}
Based on the distribution Eq. (\ref{eqn_dist}) we find the same
relation between these correlation functions, since the
reaction-plane dependent two-particle correlation is given by
\begin{eqnarray} \label{eqn_s_corr}
&& \gamma_{++/--} \equiv <\cos(\phi_i+\phi_j-2\Psi_{RP})>_{++/--}\, =d_1^2 \cos(2\,
\Delta\Psi_{CS})
\end{eqnarray}
for same-charge pairs, and
\begin{eqnarray} \label{eqn_o_corr}
&& \gamma_{+-} \equiv <\cos(\phi_i+\phi_j-2\Psi_{RP})>_{+-}\, =-d_1^2 \cos(2\,
\Delta\Psi_{CS})
\end{eqnarray}
for opposite-charge pairs.

To make contact with the predictions of the CME for the above
correlation functions, let us assume for the moment that  an accurate identification of the reaction plane could be
achieved. In this case we may rotate all events such that $\rp =
0$. Furthermore the CME predicts 
$\Delta\Psi_{CS} =\pi/2$, and thus, for $\rp=0$ the charge separation term will 
take the form of $\sim d \sin(\phi)$ 
\cite{Kharzeev:2009fn,Star:2009uh,Voloshin:2004vk}. 
If the only contribution to the above correlations would be due to the
 CME, a very specific pattern arises: 
\begin{eqnarray}
\gamma_{++/--} = <\cos(\phi_i+\phi_j-2\Psi_{RP})>_{++/--}\, &=& -\,d_1^2 < 0,
\\
\delta_{++/--} = <\cos(\phi_i-\phi_j)>_{++/--}\, &=& +\,d_1^2 > 0. 
\end{eqnarray} 
while 
\begin{eqnarray}
\gamma_{+-} = <\cos(\phi_i+\phi_j-2\Psi_{RP})>_{+-}\, &=& +\,d_1^2 > 0,
\\
\delta_{+-} = <\cos(\phi_i-\phi_j)>_{+-}\, &=& -\,d_1^2 < 0. 
\end{eqnarray} 
This pattern for the correlations $\gamma$ and $\delta$, if seen in
the data, would constitute a very strong evidence for occurrence of
the CME in these collisions. However, as pointed out in
\cite{Bzdak:2009fc},  and as we shall discuss in more detail in
Section \ref{sec:3} the STAR measurements do not show the above
pattern.  For example, while in the above
analysis for the same-charge pairs the correlators $\gamma$ 
and $\delta$ are expected to be equal in magnitude
but {\em opposite} in sign, i.e., $\gamma_{++}=-\delta_{++}$ the STAR data finds them approximately equal in magnitude but with the {\em
same} (negative) sign. 

\subsection{The $\hat{Q}_1^c$ vector analysis for measuring the charge separation}

When exploring an important  phenomenon such a local parity
violation, it is very useful to develop multiple observables which
test its predictions, such as the  Chiral 
Magnetic Effect. This is particularly the case 
in the present situation. The signals due to the CME are
expected to be rather weak and the observables are not free from various
backgrounds due to ``conventional'' physics, such as two-particle correlations. In addition, the interpretation of the STAR data is
rather ambiguous. Therefore, it will be very helpful to have  an alternative 
observable which is sensitive to a possible charge separation with specific 
azimuthal orientation. Currently there are a few proposals, for
example the  $\hat{Q}_1^c$ vector analysis 
\cite{Liao:2010nv}, the charge multiplicity asymmetry correlations \cite{Wang:2009kd}, and the out-of-plane charge asymmetry distribution \cite{Roy_group}.  Here we focus on a detailed discussion of the $\hat{Q}_1^c$ vector analysis 
\cite{Liao:2010nv}. 

The  $\hat{Q}_1^c$   vector analysis aims at a direct measurement of the
intrinsic charge-dependent distribution in Eq. (\ref{eqn_dist}) by identifying 
the charged dipole moment vector $\hat{Q}^c_1$
of the final-state hadron distribution in the transverse momentum
space. The magnitude $Q^c_1$ and azimuthal angle $\Psi^c_1$ of
this vector can be determined in a given event by the following:
\begin{eqnarray} \label{eqn_qc1_def}
Q^c_1 \cos \Psi^c_1 \equiv \sum_i q_i \cos\phi_i \nonumber \\
Q^c_1 \sin \Psi^c_1 \equiv \sum_i q_i \sin\phi_i
\end{eqnarray}
where the summation is over all charged particles in the
event, with $q_i$ the electric charge and $\phi_i$  the azimuthal angle of each particle. This method is in close
analogy to the $\hat{Q}_1$ and $\hat{Q}_2$ vector analysis used
for directed and elliptic flow (see e.g. \cite{Voloshin:2008dg}). In
the $\hat{Q}_2$ analysis one evaluates the charge independent
quadrupole moment $Q_2$ and its direction $\Psi_2$ in a similar fashion
 \begin{eqnarray} \label{eqn_q2_def}
 Q_2 \cos 2\Psi_2 \equiv \sum_i \cos2\phi_i \nonumber \\
 Q_2 \sin 2\Psi_2 \equiv \sum_i \sin2\phi_i
\end{eqnarray}
The angle $\Psi_2$ is a measure for the reaction plane angle, $\rp$ such that
for a system with infinite many particles $\Psi_2\rightarrow\rp$. 

Contrary to $\hat{Q}_2$, the charge dipole vector,
$\hat{Q}^c_1$, incorporates the {\em electric charge} $q_i$ of the
particles. The mathematical
details regarding the observable $\hat{Q^c_1}$  and its relation to multi-particle correlations 
can be found in \cite{Liao:2010nv}.
 
In each event, both angles $\Psi^c_1$ and $\Psi_2$ are  determined 
from a finite number of final state hadrons (see
Fig. \ref{fig_demo}). While these angles correspond to their
idealized expectations $\Psi_{CS}$ and $\Psi_{RP}$ only in the limit of
infinite multiplicity, their distribution and in particular
the distribution of their difference, $\Delta \Psi=\Psi^c_1- \Psi_2$
should provide a good estimator for the magnitude of the charged
dipole angle with respect to the reaction plane, $\Delta\Psi_{CS} =
\Psi_{CS}-\Psi_{RP}$.

\begin{figure}[t]
\begin{center}
\includegraphics[height=5cm]{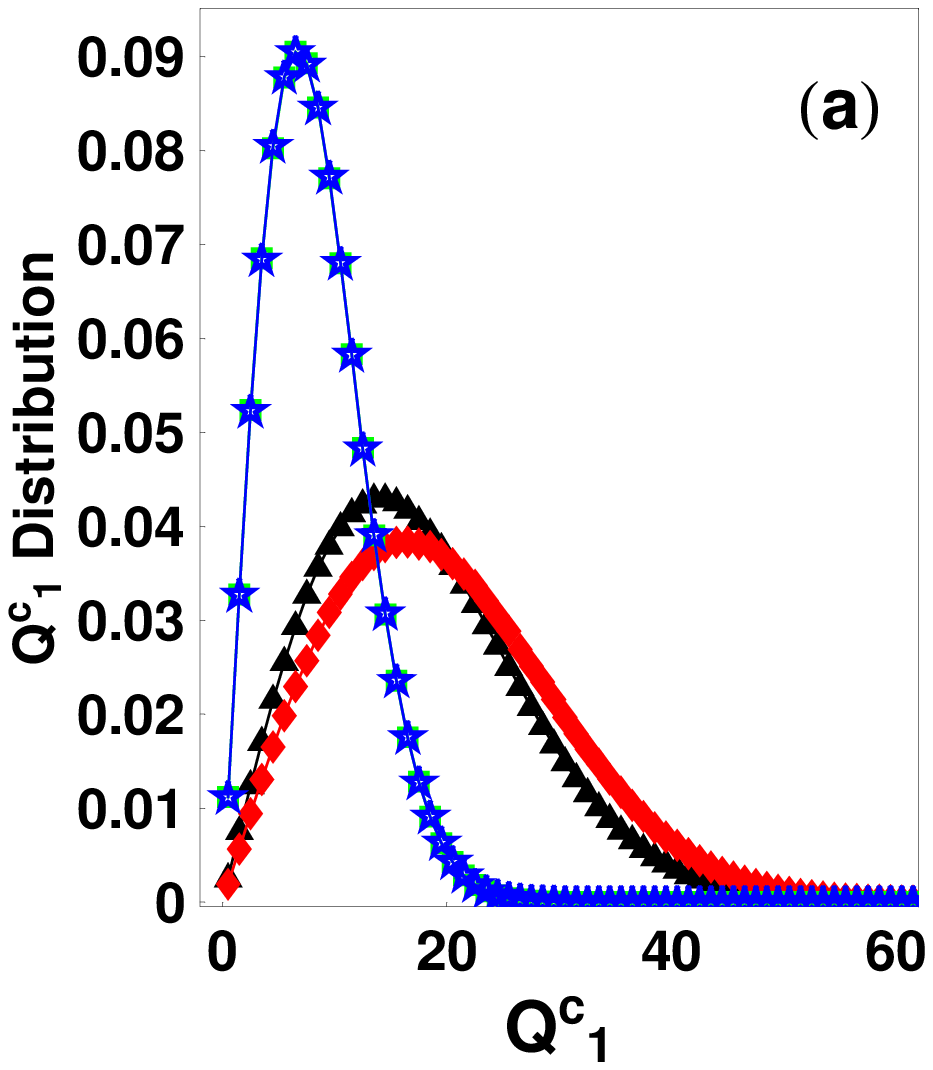}\hspace{1.cm}
\includegraphics[height=5cm]{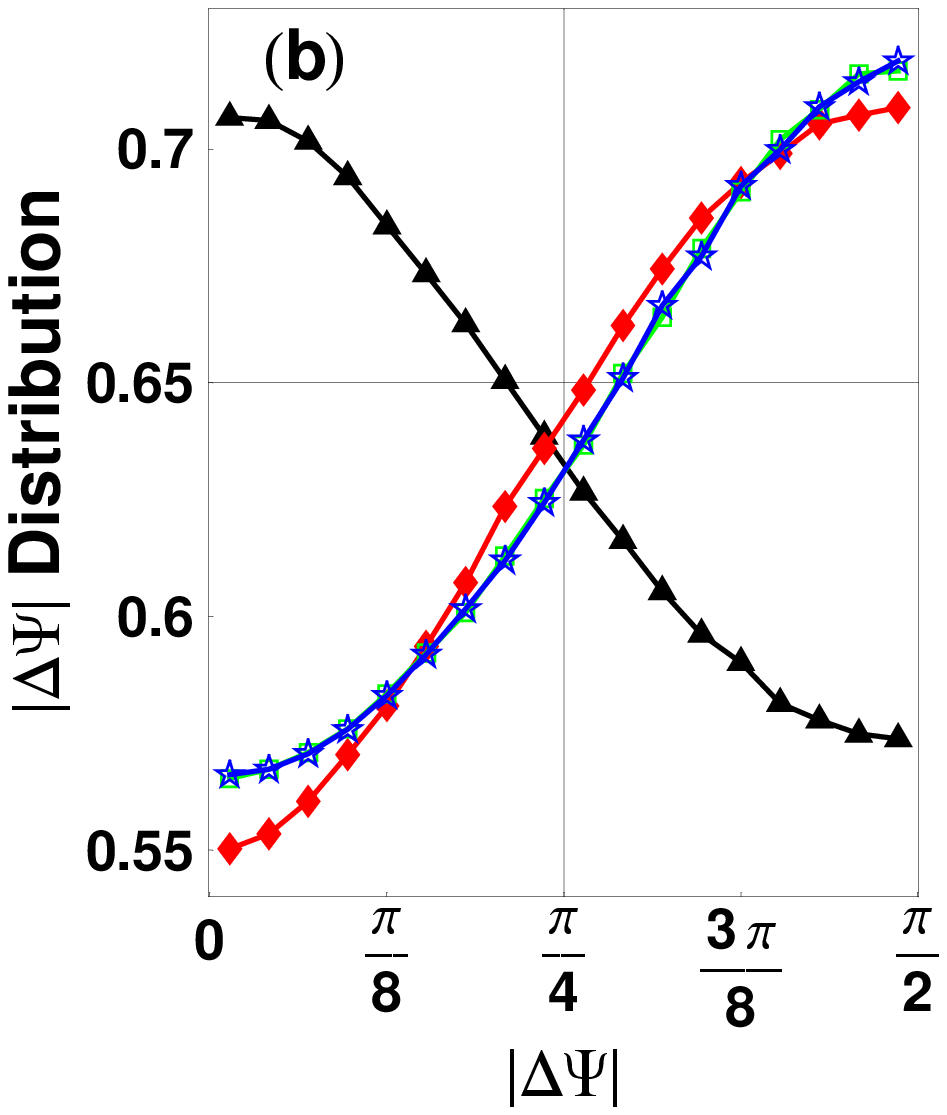}
\end{center} 
\caption{(Color online) (a) The $Q^c_1$ and (b) $|\Delta \Psi|$
distributions for the four different scenarios described in the text. }
\label{fig:dipole_analysis}
\end{figure}

The combined  $Q^c_1$- and $Q_2$- analysis will then provide
distributions for the magnitude of the electric dipole, $Q^c_1$, and
its relative angle with respect to $\Psi_2$, $\Delta \Psi$. This is
demonstrated in Fig.~\ref{fig:dipole_analysis} where we show the
distributions for various scenarios calculated in a Monte Carlo
simulation \cite{Liao:2010nv}.
\begin{itemize}
\item The black triangles correspond to a ``benchmark'' scenario,
  where we have only elliptic flow but neither 
a charged dipole nor any true pair correlations. Therefore the resulting
distributions for $Q_1^c$ and $|\Delta\Psi|$ arise only from pure statistical fluctuations.
\item The red diamonds have been obtained by adding a physical dipole
  along the out-of-plane direction with a magnitude of $d_1=0.025$, to
  the benchmark scenario.
\item The green boxes are based on the case where  back-to-back
  angular correlation for about $1\%$ of the same-charge pairs but no
  dipole have been added to the benchmark scenario. 
\item The blue stars result the case where same-side 
  angular correlation for about $1\%$ of the opposite-charge pairs but no
  dipole have been added to the benchmark scenario.  
\end{itemize}

As can be seen from the comparison in Fig.~\ref{fig:dipole_analysis}
and a more detailed  discussion in \cite{Liao:2010nv}, only the combined
analysis of the distributions of angle and magnitude, is able to
distinguish between scenarios based on conventional two-particle
correlations and those involving a true charged momentum space dipole as
predicted by the CME. As further discussed in \cite{Liao:2010nv} the
final conclusion on the possible existence of an electric dipole
will likely require a joint analysis of all
three types of measurements, discussed in this Section: the $Q^c_1$ distribution,  the 
$\Delta \Psi$ distribution, as well as the charge-dependent azimuthal correlations $\gamma$ and $\delta$.

\section{Interpretation of the available data}
\label{sec:3}

\begin{figure}[h!!]
\begin{center}
\includegraphics[height=4.5cm]{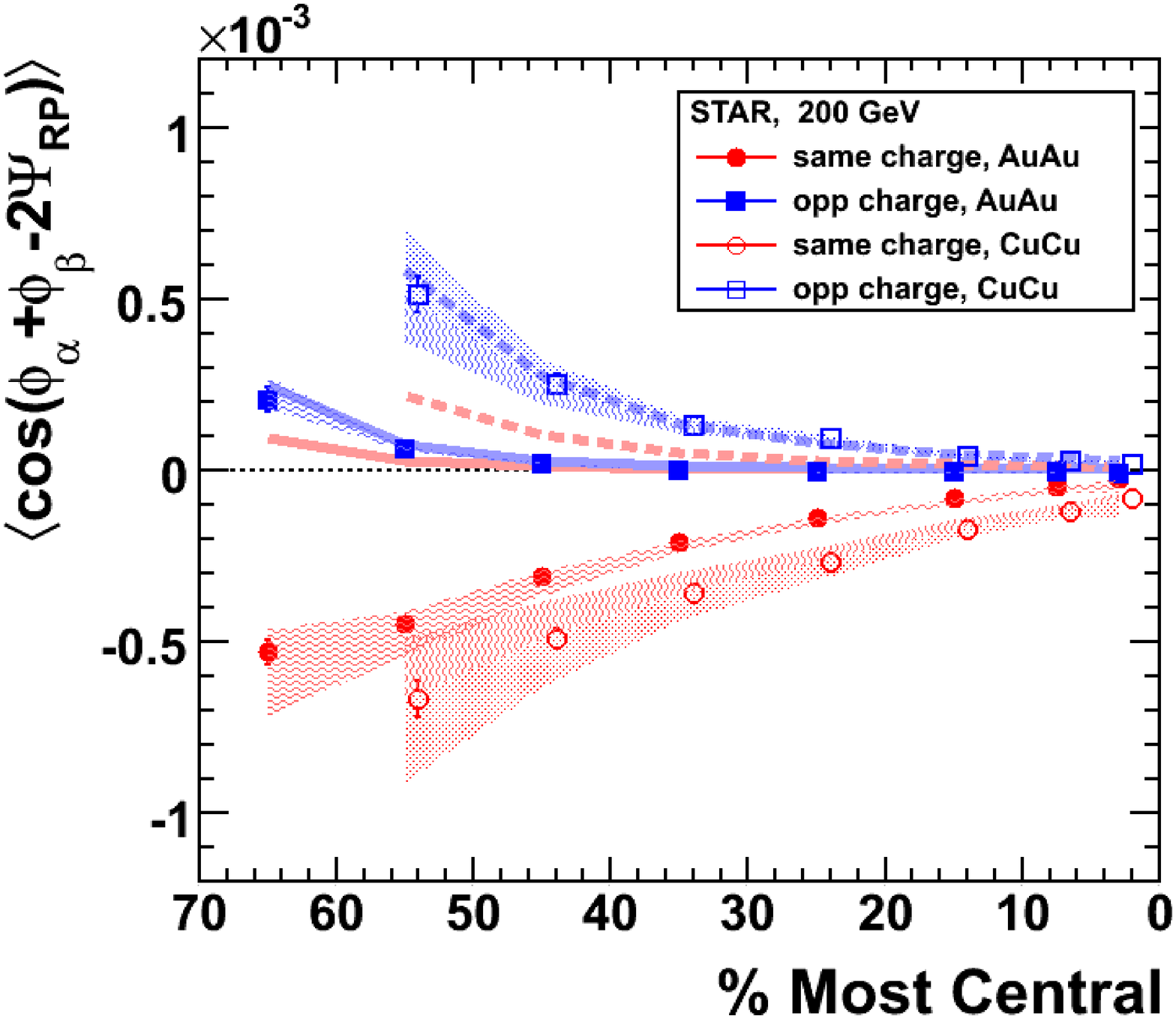}
\hspace{0.5cm}
\includegraphics[height=4.5cm]{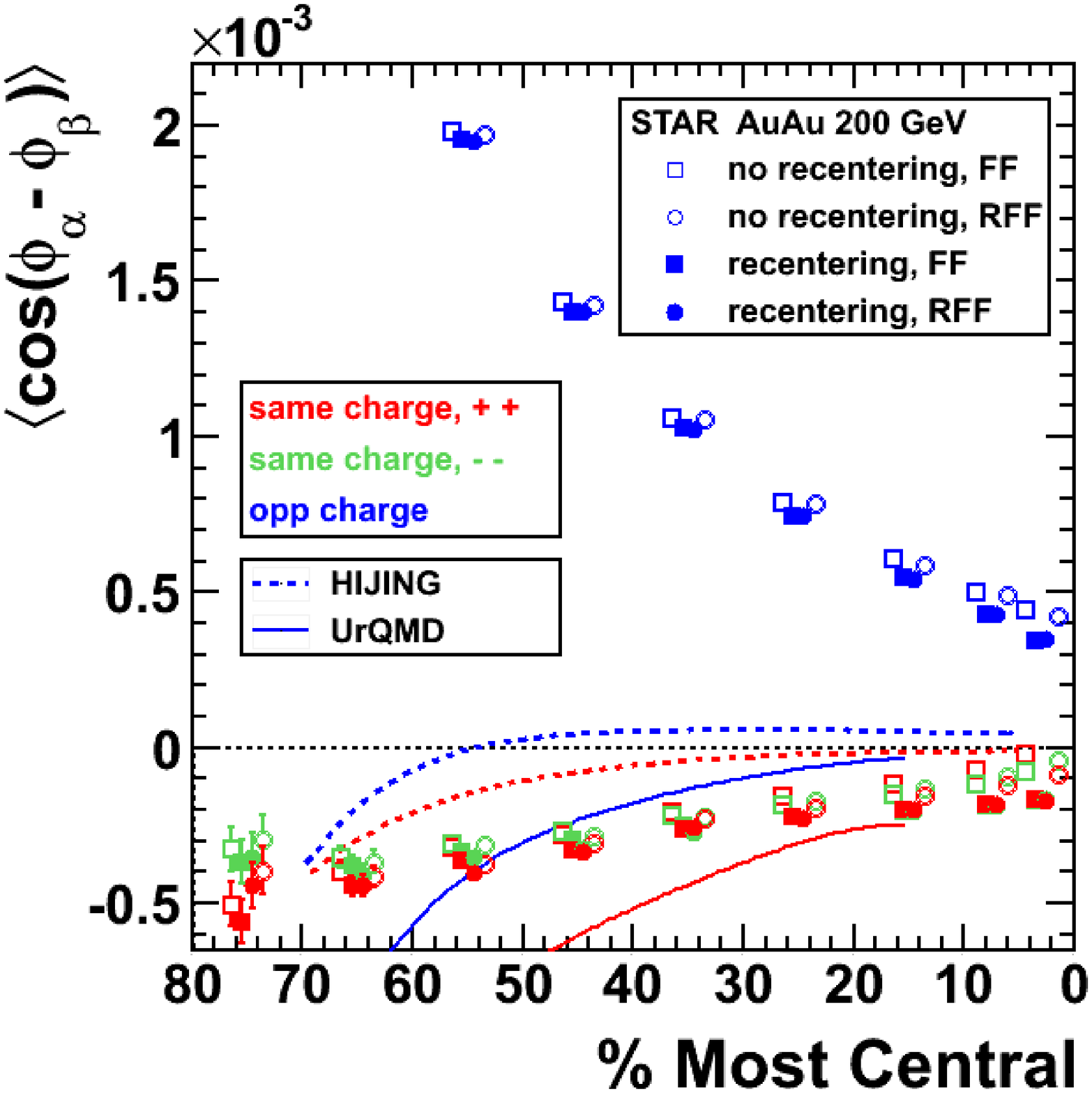}
\end{center}
\caption{The data from the STAR collaboration for the reaction plane dependent
correlation function $\ave{\cos(\phi_\alpha+\phi_\beta-2 \rp)}$ 
(left) and the reaction-plane independent correlation function
$\ave{\cos(\phi_\alpha-\phi_\beta)}$ (right) for like-sign and 
unlike-sign pairs. Also shown (lines) are  results from various 
model calculations. The Figures are from \cite{Star:2009uh}.}
\label{fig:star_data}
\end{figure}

After having discussed the general aspects of charge dependent
correlation functions in Section 2 we will now turn our
attention to the actual measurements of such correlation
function. Following the proposal by Voloshin \cite{Voloshin:2004vk}
the STAR collaboration \cite{Star:2009uh} presented the first measurement of the
reaction-plane dependent charged-pair correlation function
\be
\gamma_{\alpha,\beta}=\ave{\cos(\phi_\alpha+\phi_\beta-2 \rp)}
\label{eq:gamma_3}
\ee
for pairs of particles with same, $(\alpha,\beta)=(+,+),\,(-,-)$, and opposite
charge, $(\alpha,\beta)=(+,-)$. 
As already discussed in the previous Section, in order to obtain the
correlator $\gamma_{\alpha,\beta}$ STAR measured three-particle correlation
functions, and demonstrated rather convincingly that, within errors, they
are related to the reaction plane dependent two-particle charged--pair correlation
function by  
\be
\ave{\cos(\phi_\alpha+\phi_\beta-2\phi_k)} = v_2 \ave{\cos(\phi_\alpha+\phi_\beta-2 \rp)} =
v_2\gamma_{\alpha,\beta}
\label{eq:corr_relate}
\ee
where $v_2$ denotes the measured elliptic flow parameter characterizing the
elliptic azimuthal asymmetry. 
The results of the STAR measurement for
$\gamma_{\alpha,\beta}$ are shown in the left panel of Fig. \ref{fig:star_data}. 

Since the relation, Eq. (\ref{eq:corr_relate}), has been established
experimentally,  we will concentrate our discussion on the charge dependent
pair correlation function, $\gamma_{\alpha,\beta}$, Eq. (\ref{eq:gamma_3}). 
Furthermore, we will choose a frame where the reaction plane angle is
set to zero, $\rp=0$, so that
\be
\gamma_{\alpha,\beta}=\ave{\cos(\phi_\alpha+\phi_\beta)}.
\label{eq:gamma_simple}
\ee
In this frame the in-plane direction coincides with the x-axis and the
out-of-plane direction points along the y-axis. Also the average direction of
the magnetic field will be along the y-axis.

Before we examine the STAR data more carefully let us recall what the
prediction for the charge separation due to the Chiral Magnetic
Effects are. As discussed in the previous section, the 
electric momentum space dipole induced by the CME will point (in an ideal situation) either
parallel or anti-parallel to the direction of the magnetic field,
which in the frame where $\rp=0$ points along the y-axis (neglecting fluctuations of the magnetic field \cite{Bzdak:2011yy}). Therefore, the charge separation due to the CME predicts to have pairs of same charge preferably moving together along the positive or
negative y-direction. Pairs with opposite charge, on the other hand,
are predicted to move away from each other along the y-axis. In terms
of the azimuthal angles, $\phi_\alpha$, $\phi_\beta$ this means 
\be
(\phi_\alpha,\phi_\beta)&=&(\frac{\pi}{2},\frac{\pi}{2})\,\,{\rm or}\,\,
(\frac{3\pi}{2},\frac{3\pi}{2}) 
\ee
for same-charge pairs, and
\be
(\phi_\alpha,\phi_\beta)&=&(\frac{\pi}{2},\frac{3\pi}{2})\,\,{\rm or}\,\,
(\frac{3\pi}{2},\frac{\pi}{2}) 
\ee 
for opposite-charge pairs. Since 
\be
\cos(\frac{\pi}{2}+\frac{\pi}{2})&=&\cos(\frac{3\pi}{2}+\frac{3\pi}{2})=-1
\\
\cos(\frac{3\pi}{2}+\frac{\pi}{2})&=&1
\ee
the correlation function $\gamma_{\alpha,\beta}$, Eq. (\ref{eq:gamma_simple}), is
expected to be negative for same-charge pairs and positive for
opposite-charge pairs. While the STAR data, shown in
Fig.~\ref{fig:star_data}, indeed show a negative value for same-charge pairs,
the result for opposite-charge pairs is, at best, only mildly
positive and, within errors, compatible with zero. Since opposite
charged pairs are predicted to move away from each other, one may
argue their (anti-) correlation should be weakened as these particles
will have to traverse the entire fireball \cite{Kharzeev:2007jp}. 
Therefore, at first sight,
the STAR data may indeed show a first evidence for the charge
separation pattern as predicted by the CME. However, the
interpretation of the data is more difficult.

The complication arises from the fact that the correlation function
$\gamma_{\alpha,\beta}$ does not unambiguously determine the
angular correlation of the pair. To see this, consider a same-charge
pair with angles $(\phi_\alpha,\phi_\beta)=(0,\pi)$. In this case the
particles move away from each other in the in-plane direction. This is just 
the opposite of the correlation predicted by the CME,
where the two particles are moving with each other in the out-of-plane direction. 
For both cases we get
\be
\cos(\phi_\alpha,\phi_\beta)=\cos(0+\pi)=\cos(\frac{\pi}{2}+\frac{\pi}{2})=-1. 
\ee
Thus, the correlation function $\gamma_{\alpha,\beta}$ is {\em not} able
to distinguish between same-side out-of-plane correlations and
back-to-back in-plane correlations. However, this ambiguity can easily
be resolved by considering the reaction plane independent correlation
function
\be
\delta_{\alpha,\beta} = \ave{\cos(\phi_\alpha-\phi_\beta)}
\label{delta_3}
\ee
which STAR has also measured, and we show their results in the right
panel of Fig.~\ref{fig:star_data}. In the frame, where $\rp=0$, the
two correlation functions may be decomposed in the in-plane 
$\sim\ave{\cos(\phi_\alpha)\cos(\phi_\beta)}$ and out-of-plane
$\sim\ave{\sin(\phi_\alpha)\sin(\phi_\beta)}$ components:   
\begin{eqnarray}
\gamma_{\alpha,\beta}=\left\langle \cos (\phi _{\alpha }+\phi _{\beta })\right\rangle
&=&\left\langle \cos (\phi _{\alpha })\cos (\phi _{\beta })\right\rangle
-\left\langle \sin (\phi _{\alpha })\sin (\phi _{\beta })\right\rangle , 
\non 
\delta_{\alpha,\beta}=\left\langle \cos (\phi _{\alpha }-\phi _{\beta })\right\rangle
&=&\left\langle \cos (\phi _{\alpha })\cos (\phi _{\beta })\right\rangle
+\left\langle \sin (\phi _{\alpha })\sin (\phi _{\beta })\right\rangle .
\label{math}
\end{eqnarray}

Qualitatively the STAR measurement in Au+Au collisions for both these correlation
functions, $\gamma_{\alpha,\beta}$ and $\delta_{\alpha,\beta}$ for same-sign 
and opposite-sign pairs of charged particles, may be
characterized as follows (see Fig.~\ref{fig:star_data}): 
\begin{itemize}
\item For same-sign pairs:%
\begin{equation}
\left\langle \cos (\phi _{\alpha }+\phi _{\beta })\right\rangle
_{same}\simeq \left\langle \cos (\phi _{\alpha }-\phi _{\beta
})\right\rangle _{same}<0.
\end{equation}%
Using Eq. (\ref{math}) this implies%
\begin{eqnarray}
\left\langle \sin (\phi _{\alpha })\sin (\phi _{\beta })\right\rangle
_{same} &\simeq &0,  \non 
\left\langle \cos (\phi _{\alpha })\cos (\phi _{\beta })\right\rangle
_{same} &<&0.  \label{ss0}
\end{eqnarray}

\item For opposite-sign pairs we find that%
\begin{eqnarray}
\left\langle \cos (\phi _{\alpha }+\phi _{\beta })\right\rangle _{opposite}
&\simeq &0  \non 
\left\langle \cos (\phi _{\alpha }-\phi _{\beta })\right\rangle _{opposite}
&>&0.
\end{eqnarray}

Again, using Eq. (\ref{math}), this means%
\begin{equation}
\left\langle \sin (\phi _{\alpha })\sin (\phi _{\beta })\right\rangle
_{opposite}\simeq \left\langle \cos (\phi _{\alpha })\cos (\phi _{\beta
})\right\rangle _{opposite}>0.
\end{equation}

\end{itemize}

\begin{figure}[h!!]
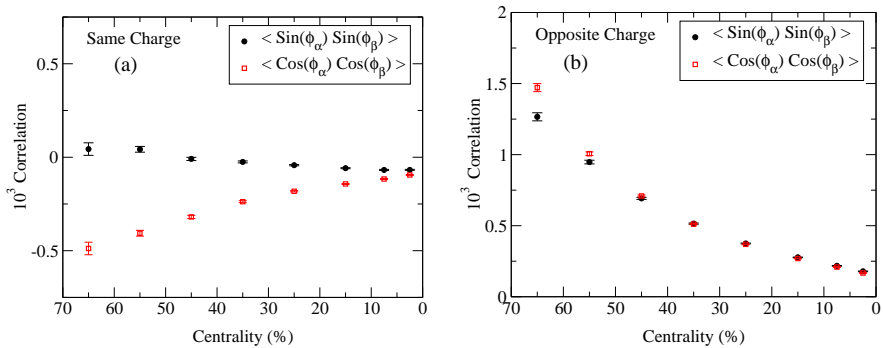

\begin{center}
\includegraphics[width=0.48\textwidth]{same.eps}
\hspace{0.2cm}
\includegraphics[width=0.48\textwidth]{opposite.eps}
\end{center}
\caption{Correlations in-plane $\left\langle \cos (\protect\phi _{\protect%
\alpha })\cos (\protect\phi _{\protect\beta })\right\rangle $ and out-of-plane 
$\left\langle \sin (\protect\phi _{\protect\alpha })\sin (\protect\phi %
_{\protect\beta })\right\rangle $ for same- and opposite-charge pairs in $Au+Au$ collisions as seen in the STAR data.}
\label{fig_sscc}
\end{figure}

\begin{figure}[h]
\begin{center}
\includegraphics[width=0.4\textwidth]{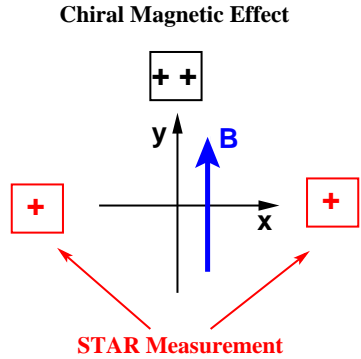} 
\end{center}
\caption{Schematic illustration of the actual STAR measurement (red)
together with the predictions from the Chiral Magnetic Effect (black) 
for same-charge pairs.}
\label{fig:star_cartoon}       % Give a unique label
\end{figure}

The decomposition of the actual data into the in-plane and
out-of-plane components is shown in Fig. \ref{fig_sscc}.
Obviously the correlations for same-charge pairs are predominantly  in-plane
and back-to-back. This is exactly the {\em opposite} of what has been
predicted by the Chiral Magnetic Effect. This is illustrated in  
Fig. \ref{fig:star_cartoon} were we have sketched the experimental
situation for same-charge pairs based on the STAR data. For pairs
with opposite charge, both in-plane and out-of-plane correlations have
the same (positive) sign and magnitude. This implies that
opposite-charged pairs move together equally likely in the in-plane
and out-of-plane directions. This behavior can at least qualitatively
be understood by resonance/cluster decays \cite{Wang:2009kd} or local charge
conservation \cite{Schlichting:2010qia}.

In addition to the data shown in Fig.~\ref{fig:star_data}, STAR has
also analyzed the reaction plane dependent correlation function
$\gamma_{\alpha,\beta}$ differentially as a function of the pair
transverse momentum (sum and difference) and rapidity difference. Both these results are within qualitative expectations for a charge separation effect due
to the CME \cite{Bzdak:2009fc}. Unfortunately, similar differential
information is not available for the reaction plane independent
correlation function, $\delta_{\alpha,\beta}$. Therefore a
differential decomposition into in-plane and out-of-plane components,
as we have done here, unfortunately is not possible at this time. Such
information may help to further constrain possible background effects as
well as predictions from the CME.

Recently, the ALICE collaboration reported \cite{Abelev:2012pa} the measurement of the same
correlation functions for Pb+Pb collisions at a center of mass energy
of $\sqrt{s}=2.76 \, {\rm TeV}$, about ten times that of the STAR
measurement. Just like STAR, ALICE determined the reaction plane
dependent correlation function $\gamma_{\alpha,\beta}$ integrated 
over transverse momentum and rapidity as well as differentially.  
Within errors, the data for the integrated correlation function
$\gamma_{\alpha,\beta}$ agree with those of
the STAR measurement, and  the differential measurement show the same
qualitative features. 

For the reaction plane independent correlation
function, $\delta_{\alpha,\beta}$, on the other hand, the ALICE date
differ from those by STAR. In particular, ALICE finds this correlation
function to be positive for {\em both} opposite- and same-charge
pairs.  ALICE also provides the  in-plane and
out-of-plane pair correlations, $\ave{\cos(\phi_\alpha)\cos(\phi_\beta)}$ and
$\ave{\sin(\phi_\alpha)\sin(\phi_\beta)}$, respectively. Similar to the
STAR measurement ALICE finds that for opposite-charge pairs the in-
and out-of-plane correlations are nearly identical and positive. For
the same-charge pairs, however, ALICE finds both in- and out-of-plane
projections to be positive, with the out-of-plane correlation slightly
larger than the in-plane projection. This finding would be in
qualitative agreement with the expectations from the CME. Amusingly,
early predictions \cite{Toneev:2010xt} for the collision energy
dependence of the CME expected a smaller effect at the very high
energies where ALICE has 
been measuring, largely due to the shorter duration of the magnetic
field. Of course the complex dynamics of heavy ion collisions and the various
background contributions turn quantitative predictions for these
correlation functions into a very difficult task, and a final
resolution will 
require a systematic analysis of all available data at various energies. 

Given that the STAR data show an in-plane back-to-back correlation for
same-sign pairs, one may wonder if there is still room for a charge
separation effect due to the CME. This has been analyzed in
\cite{Bzdak:2009fc} with the result that for the transverse momentum
and rapidity integrated data, which the above analysis is based on, the
backgrounds need to exactly cancel the CME induced charge
separation. This may be just a coincidence, however. After all the data by
ALICE show a different trend for the same-sign correlations. For this
situation to be clarified further differential data for the reaction
plane independent correlation function, $\delta_{\alpha,\beta}$ are
required for both collision energies. 

Finally, as part of the RHIC beam energy scan program, STAR has
measured the reaction-plane dependent correlator
$\gamma_{\alpha,\beta}$ for various collision energies
\cite{Mohanty:2011nm}, see also \cite{Toneev:2011aa}. They find that the difference for the
correlator between same-sign and opposite-sign pairs decreases with
decreasing beam energy. Such a behavior is expected from the 
CME. However, as we will discuss in the next section, all background terms 
scale with the elliptic flow parameter, $v_2$, which is known to
decrease with decreasing collision energy as well.

To conclude this section, presently available experimental results
concerning the CME are inconclusive. While the integrated STAR data
disfavor the presence of the CME, the ALICE data allow for more
positive conclusions. Clearly, progress requires data at lower energies
as well as, and most importantly, differential measurements of both
the reaction plane dependent and reaction plane independent
correlation functions. In addition, given the rather unsettled state
of affairs, measurements of other observables, such as the one
proposed in the previous section, would be very welcome.

\section{Discussion of various background contributions}
\label{sec:4}

As discussed in the previous sections the Chiral Magnetic Effect (CME)
- if it exists - contributes to the reaction plane dependent 
two-particle correlator, first introduced in \cite{Voloshin:2004vk}. As in
the previous section we denote the reaction plane dependent two-particle 
correlator by%
\begin{equation}
\gamma \equiv \left\langle \cos (\phi _{1}+\phi _{2}-2\Psi
_{RP})\right\rangle ,  \label{gam-def}
\end{equation}%
where $\phi _{1}$ and $\phi _{2}$ are the azimuthal angles of two particles,
and $\Psi_{RP}$ is the reaction plane angle. In the following we will distinguish between $%
\gamma _{++/--}$, $\gamma _{+-}$ and $\gamma $ denoting respectively the
correlator (\ref{gam-def}) for same-sign pairs, opposite-sign pairs, and the
correlator without specifying the sign of measured particles.

As discussed in Section \ref{sec:3} a detailed measurement of $\gamma $ was performed both at RHIC by the STAR collaboration \cite{Star:2009uh} and at the LHC by the ALICE
collaboration \cite{Abelev:2012pa}.
However, as already discussed in the previous Section, the interpretation of experimental data is not straightforward
since various effects can contribute to $\gamma $. 

The presence of elliptic flow allows for practically all ``conventional'' two-particle correlations to contribute to the reaction-plane dependent correlation function, $\gamma$.
This can be easily seen from the decomposition of $\gamma$ into in-plane and out-of-plane
projections, Eq. (\ref{math})
\begin{equation}
\gamma =\left\langle \cos (\phi _{1})\cos (\phi _{2})\right\rangle
-\left\langle \sin (\phi _{1})\sin (\phi _{2})\right\rangle .
\end{equation}%
It is quite obvious that even if the underlying correlation mechanism does not
depend on the reaction plane it will contribute to $\gamma $ in the presence
of the elliptic anisotropy $v_{2}$. This can be seen in an extreme,
though  unrealistic, situation where all particles are produced exactly in-plane. In this case 
$\left\langle \sin(\phi _{1})\sin (\phi _{2})\right\rangle =0$ simply because there are no
particles in the out-of-plane direction and 
$\gamma = \left\langle \cos (\phi _{1})\cos (\phi _{2})\right\rangle$. Obviously in this case, the presence of any two-particle angular correlation mechanism will result in a non-zero value of $\gamma$.

In this Section we will focus exclusively on the contribution to
$\gamma $ driven by the non-vanishing elliptic anisotropy $v_{2}$.
First we will derive the general expression which
relates the elliptic anisotropy and the correlator $\gamma $ in the presence
of  arbitrary two-particle correlations. Next we will discuss a few explicit
mechanisms that need to be understood quantitatively,  before any
conclusions about the existence of the CME can be made.
In particular we will address corrections due to
transverse momentum conservation (TMC) \cite{Pratt:2010zn,Bzdak:2010fd,Ma:2011um} 
and the local charge conservation \cite{Schlichting:2010qia}, both of which appear to contribute significantly to $\gamma $.
In the last part of the paper we will discuss the possibility of removing,
in the model independent way, the elliptic-flow-related background from $\gamma $. 

\subsection{General relation}

In this part  we derive the general relation between the elliptic
anisotropy $v_{2}$ and the two-particle correlator $\gamma $ in the presence
of an arbitrary reaction plane independent two-particle correlations.

By definition, the two-particle correlator $\gamma $ is 
\begin{equation}
\gamma =\frac{\int \rho _{2}(\phi _{1},\phi _{2},x_{1},x_{2},\Psi _{RP})\cos
(\phi _{1}+\phi _{2}-2\Psi _{RP})d\phi _{1}d\phi _{2}dx_{1}dx_{2}}{\int \rho
_{2}(\phi _{1},\phi _{2},x_{1},x_{2},\Psi _{RP})d\phi _{1}d\phi
_{2}dx_{1}dx_{2}},  \label{gam-i}
\end{equation}%
where, to simplify our notation, we denote: $x=(p_{t},\eta )$ and $%
dx=p_{t}dp_{t}d\eta $. Here $p_{t}$ is the absolute value of
transverse-momentum, while $\eta $ is pseudorapidity (or rapidity). $\rho
_{2}$ is the two-particle distribution in the intrinsic frame with the
reaction plane angle  $\Psi _{RP}$. It can be expressed in terms of
the single-particle distributions, and the
underlying correlation function $C$ (see Section~\ref{sec:2})
\begin{equation}
\rho _{2}(\phi _{1},\phi _{2},x_{1},x_{2},\Psi _{RP})=\rho (\phi
_{1},x_{1},\Psi _{RP})\rho (\phi _{2},x_{2},\Psi _{RP})[1+C(\phi _{1},\phi
_{2},x_{1},x_{2})].  \label{ro2}
\end{equation}%
To simplify our calculation we assume the single-particle distribution
to be
\begin{equation}
\rho (\phi ,x,\Psi _{RP})=\frac{\rho _{0}(x)}{2\pi }[1+2v_{2}(x)\cos \left(
2\phi -2\Psi _{RP}\right) ],  \label{ro1}
\end{equation}%
where $\rho _{0}(x)$ and $v_2(x)$ depend solely on $x=(p_{t},\eta
)$. We neglect higher moments $v_{n}$ since their contribution to $\gamma $ turns out to be
proportional to $v_{n}v_{m}$ which is much smaller then the leading
term $\sim v_{2}$, see Ref. \cite{Bzdak:2011my}.

If $v_{2}(x)\neq 0$, the single particle distributions depend on the reaction 
plane. Therefore, the part of the
two-particle density (\ref{ro2}) involving the two-particle
correlation function $C$ depends on the reaction plane even if 
$C$ depends only on $\phi_{1}-\phi_{2}$.

Here we want to concentrate on
those correlations that depend only on $\Delta \phi =\phi _{1}-\phi _{2}$,
namely the underlying correlation mechanism is insensitive to the reaction
plane orientation. The correlation function may be expanded in a Fourier
series%
\begin{equation}
C(\Delta \phi ,x_{1},x_{2})=\sum\nolimits_{n=0}^{\infty
}a_{n}(x_{1},x_{2})\cos \left( n\Delta \phi \right) ,  \label{c-fourier}
\end{equation}%
where $a_{n}(x_{1},x_{2})$ does not depend on $\phi _{1}$ and $\phi _{2}$.
Substituting (\ref{c-fourier}) and (\ref{ro2}) into Eq. (\ref{gam-i}), we
obtain%
\begin{equation}
\gamma =\frac{1}{2N^{2}}\int \rho _{0}(x_{1})\rho
_{0}(x_{2})a_{1}(x_{1},x_{2})[v_{2}(x_{1})+v_{2}(x_{2})]dx_{1}dx_{2},
\label{gam-v2}
\end{equation}%
where $N=\int \rho _{0}(x)dx$, and we have assumed that
$a_{n}<<1$.

Equation (\ref{gam-v2}) explains why all correlation mechanisms with a
non-zero $a_{1}(x_{1},x_{2})$ contribute to $\gamma $. For instance, 
it has been shown that transverse momentum conservation (TMC)
leads to a correlation function which depends on $\cos
\left( \Delta \phi \right) /N_{\mathrm{tot}}$ \cite{Borghini:2000cm}, where $%
N_{\mathrm{tot}}$ is the total number of produced particles. In this case $%
a_{1}(x_{1},x_{2})\propto 1/N_{\mathrm{tot}}$. Let us also emphasize that
all correlations that depend on the momentum difference between particles $%
\Delta k=|\vec{k}_{1}-\vec{k}_{2}|$ also contribute to $\gamma $. In this
case:%
\begin{equation}
C(\Delta k)=C\left( k_{1}^{2}+k_{2}^{2}-2k_{1}k_{2}\cos (\Delta \phi
)\right) ,  \label{c-dk}
\end{equation}%
which naturally leads to a non-vanishing $a_{1}$ term.

To summarize, Eq. (\ref{gam-v2}) explains why transverse-momentum
conservation \cite{Pratt:2010zn,Bzdak:2010fd,Ma:2011um}, local
charge-conservation \cite{Schlichting:2010qia}, resonance- (cluster-) decay %
\cite{Wang:2009kd}, and all other correlations with $\Delta \phi $
dependence contribute to $\gamma $.

\subsection{Transverse momentum conservation}

Very soon after publication of the experimental data by the STAR
Collaboration it was realized that transverse momentum conservation
convoluted with the non-zero elliptic anisotropy can lead to a substantial
corrections for $\gamma $
\cite{Pratt:2010gy,Bzdak:2010fd,Pratt:2010zn}. This can  be
easily seen for the simplified situation where all particles are measured (in
the full phase-space), and where they all have exactly the same magnitude of
transverse momentum $|\vec{p}_{i,t}|=|\vec{p}_{t}|$, $i=1,...,N_{\mathrm{tot}}$.

In the frame where $\rp=0$ the correlator $\gamma $ may be
written as 
\begin{equation}
\gamma =\left\langle \frac{\sum\nolimits_{i\neq j}\cos (\phi _{i}+\phi _{j})%
}{\sum_{i\neq j}1}\right\rangle ,
\end{equation}%
or, alternatively,%
\begin{equation}
\gamma =\left\langle \frac{\left( \sum\nolimits_{i}\cos (\phi _{i})\right)
^{2}-\left( \sum\nolimits_{i}\sin (\phi _{i})\right)
^{2}-\sum\nolimits_{i}\cos (2\phi _{i})}{\sum_{i\neq j}1}\right\rangle ,
\label{gam_dif}
\end{equation}%
where $i$ and $j$ are summed over all particles in the full phase-space. In
the simplified scenario, where $|\vec{p}_{i,t}|=|\vec{p}_{t}|$, the
conservation of transverse momentum implies 
\begin{equation}
\sum\nolimits_{i}\cos (\phi _{i})=\sum\nolimits_{i}\sin (\phi _{i})=0.
\end{equation}%
Consequently we obtain 
\begin{equation}
\gamma =-\left\langle \frac{\sum\nolimits_{i}\cos (2\phi _{i})}{N_{\mathrm{%
tot}}(N_{\mathrm{tot}}-1)}\right\rangle \approx \frac{-v_{2}}{N_{\mathrm{tot}%
}},  \label{gam-sp}
\end{equation}%
where $N_{\mathrm{tot}}$ is the total number of particles. Taking, for
example, the
centrality class $40-50\%$ we approximately have $v_{2}\approx 0.1$ and $%
N_{\mathrm{tot}}\approx 1500$ leading to $\gamma \approx -0.7\cdot
10^{-4} $ from TMC. This is roughly a factor $3-4$ smaller than
the experimental data for the same-charge pairs. This is only a simple estimation and a more realistic AMPT calculations \cite{Ma:2011um} suggest that the TMC contribution is roughly factor 2 smaller than the STAR data.   

In a similar way we obtain for the reaction plane independent
correlation function%
\begin{equation}
\delta =\left\langle \cos (\phi _{1}-\phi _{2})\right\rangle \approx -\frac{1%
}{N_{\mathrm{tot}}}.
\end{equation}%
In this case for $N_{\mathrm{tot}}\approx 1500$ we obtain $\delta \approx
-0.7\cdot 10^{-3}$ which is comparable or slightly larger in
magnitude than same-charge data by the STAR collaboration.

Similar results hold also in a more realistic situation, where
only a small fraction of all particles is measured, and the magnitudes of
transverse momenta are distributed according to the thermal
distribution. This has been discussed in detail in
\cite{Bzdak:2010fd}, and we will only show the most important results. 

Using the central limit theorem and implying the global conservation of
transverse momentum, the two-particle distribution function reads \cite%
{Borghini:2000cm,Chajecki:2008vg,Bzdak:2010fd} 
\begin{equation}
\rho _{2}(\vec{p}_{1},\vec{p}_{2})\simeq \rho (\vec{p}_{1})\rho (\vec{p}%
_{2})\left( 1+\frac{2}{N_{\mathrm{tot}}}-\frac{(p_{1,x}+p_{2,x})^{2}}{2N_{%
\mathrm{tot}}\left\langle p_{x}^{2}\right\rangle _{F}}-\frac{%
(p_{1,y}+p_{2,y})^{2}}{2N_{\mathrm{tot}}\left\langle p_{y}^{2}\right\rangle
_{F}}\right) .  \label{ro2-tmc}
\end{equation}%
where $x$ and $y$ denote the two components of transverse momentum. $F$
denotes that the appropriate average is calculated for all particles in the
full phase-space. The single particle distribution, $\rho (\vec{p}_{1})$, is 
given by Eq. (\ref{ro1}).

Before we continue  let us clarify one subtle point. Equation (\ref{ro2-tmc})
is derived assuming that  we first sample particles with a given $v_{2}$ and
next we conserve transverse momentum for all particles. In reality  the
opposite scenario should be considered. First we should sample
partons/particles with a conserved transverse momentum, and after this the
elliptic anisotropy $v_{2}$ should be generated according to some dynamical
model. Of course the second approach is much more challenging and renders
analytical calculations difficult. At the end of this  Section 
we will show that both procedures lead to comparable results for 
$\gamma $ and $\delta $ and their transverse momentum 
distributions, however the rapidity distributions are quite different. 
We will come back to this point later.

Using Eq. (\ref{ro2-tmc}) and Eq. (\ref{gam-i}) we can derive the following
relations for $\gamma $ and $\delta $: 
\begin{equation}
\gamma =-\frac{1}{N_{\mathrm{tot}}}\frac{\left\langle p_{t}\right\rangle
_{\Omega }^{2}}{\left\langle p_{t}^{2}\right\rangle _{F}}\frac{2\bar{v}%
_{2,\Omega }-\bar{\bar{v}}_{2,F}-\bar{\bar{v}}_{2,F}(\bar{v}_{2,\Omega })^{2}%
}{1-\left( \bar{\bar{v}}_{2,F}\right) ^{2}},
\end{equation}%
and 
\begin{equation}
\delta =-\frac{1}{N_{\mathrm{tot}}}\,\frac{\left\langle p_{t}\right\rangle
_{\Omega }^{2}}{\left\langle p_{t}^{2}\right\rangle _{F}}\,\frac{1+(\bar{v}%
_{2,\Omega })^{2}-2\bar{\bar{v}}_{2,F}\,\bar{v}_{2,\Omega }}{1-\left( \bar{%
\bar{v}}_{2,F}\right) ^{2}},  \label{eq_minus}
\end{equation}%
where we have introduced certain weighted moments of $v_{2}$:%
\begin{equation}
\bar{v}_{2}=\frac{\left\langle v_{2}(p_{t},\eta )p_{t}\right\rangle }{%
\left\langle p_{t}\right\rangle },\quad \bar{\bar{v}}_{2}=\frac{\left\langle
v_{2}(p_{t},\eta )p_{t}^{2}\right\rangle }{\left\langle
p_{t}^{2}\right\rangle }.  \label{v2-=}
\end{equation}%
In the above equations $F$ and $\Omega$ denote averages that are
calculated for all particles in the full phase-space, or for all
actually measured particles in the restricted phase-space, respectively.  
Performing explicit calculations with reasonable assumptions about $p_{t}$
and $\eta $ dependence of the single-particle distribution $\rho (p_{t},\eta
)$ and elliptic flow $v_{2}(p_{t},\eta )$, we have found that for
mid-central and peripheral collisions \cite{Bzdak:2010fd}%
\begin{equation}
\gamma \cdot N_{part}\approx -0.005,\quad \delta \cdot N_{part}\approx -0.05,
\end{equation}%
where $N_{part}$ is the number of participants, also referred to as 
wounded nucleons~\cite{Bialas:1976ed}.

To summarize, transverse momentum conservation results in a negative contributions
to both $\gamma $ and $\delta $, and they are of the same order of magnitude as the
experimental measurement for like-sign pairs. More precisely they are a factor of $3-5$ (very
peripheral -- mid-central) less in magnitude for $\gamma $, and a factor 
$1.5-4$ (mid-central -- very peripheral) larger for $\delta $ than the STAR
data for the same-sign correlator. While it is rather difficult to
understand the data with only transverse momentum conservation it is
interesting to notice that the STAR experiment is sensitive enough to the
effect of global transverse momentum conservation.

Using Eqs. (\ref{gam-i},\ref{ro2-tmc}) we can easily calculate the
dependence of $\gamma $ and $\delta $ on the sum $%
p_{+}=(p_{1,t}+p_{2,t})/2$ and difference $p_{-}=|p_{1,t}-p_{2,t}|/2$ 
of the transverse momenta of the pair. For the STAR data 
\cite{Star:2009uh}, $\gamma $ is growing roughly linearly with $p_{+}$ and is
approximately constant as a function of $p_{-}$. 
\begin{figure}[h!!]
\begin{center}
\includegraphics[scale=0.6]{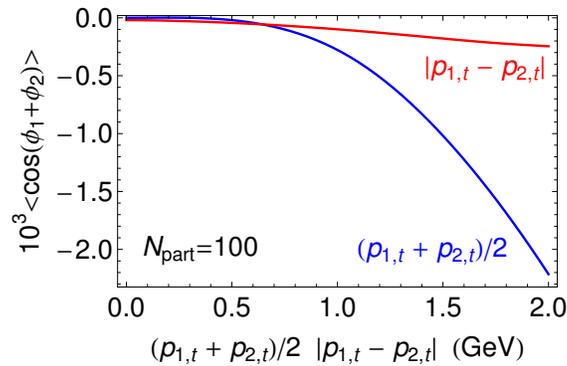}
\end{center}
\caption{The two-particle azimuthal correlator $\left\langle \cos (\protect%
\phi _{1}+\protect\phi _{2})\right\rangle $ vs $p_{+}=(p_{1,t}+p_{2,t})/2$
(blue line) and $p_{-}=\left| p_{1,t}-p_{2,t}\right| $ (red line) for $%
N_{part}=100$. The results are in qualitative and partly quantitative
agreement with the STAR data for the same-charge correlator. Figure
from Ref.~\cite{Bzdak:2010fd}.}
\label{ab_fig1}
\end{figure}
Interestingly a very similar
behavior is obtained in the scenario with only global transverse momentum
conservation. As seen in Fig. \ref{ab_fig1}, the contribution of TMC to $%
\gamma $ is consistent with the data for $p_{+}>1$ GeV and underestimates
the data for $p_{+}<1$ GeV. As expected, for $\delta $ very similar
dependence on $p_{+}$ and $p_{-}$ is obtained but rescaled by a value of $%
v_{2}$. 

Recently, very similar results were obtained in the AMPT model calculation %
\cite{Ma:2011um}, where the transverse momentum is conserved on the
event-by-event basis, and we will come back to this point later.

\subsubsection{Pseudorapidity dependence}

Since the contribution to $\gamma$ due to transverse momentum conservation 
is proportional to $v_2$, one would
naively expect that its (pseudo)rapidity dependence trace the rather
mild (pseud)rapidity dependence of $v_2$.

However, as shown in Ref. \cite{Pratt:2010zn}, under quite reasonable 
assumptions we can obtain very
similar rapidity dependence as in the experimental data. In the STAR
measurement \cite{Star:2009uh} the correlator $\gamma $ is maximum 
for $|\eta _{1}-\eta _{2}|=0$
and is approximately linearly decreasing to values consistent with zero at 
$|\eta _{1}-\eta _{2}|\approx 2$.

For simplicity of the argument let us assume that at the time $t=0$  all
produced partons/particles are distributed between two separate bins in
rapidity with enforced global transverse momentum conservation. It means
that if in the first bin the total transverse momentum equals $\vec{K}_{1,t}$%
, in the second bin $\vec{K}_{2,t}=-$ $\vec{K}_{1,t}$ in each event.
Assuming further that all particles have the same magnitude of transverse
momentum $|\vec{p}_{i,t}|=|\vec{p}_{t}|$, $i=1,...,N_{\mathrm{tot}}$, we
obtain the following relation%
\begin{equation}
\gamma \propto \left\langle \sum\limits_{i\in 1}\cos (\phi
_{i})\sum\limits_{k\in 2}\cos (\phi _{k})-\sum\limits_{i\in 1}\sin (\phi
_{i})\sum\limits_{k\in 2}\sin (\phi _{k})\right\rangle ,
\end{equation}%
and 
\begin{equation}
\gamma \propto \left\langle K_{x,1}K_{x,2}-K_{y,1}K_{y,2}\right\rangle
=\left\langle K_{y,1}^{2}-K_{x,1}^{2}\right\rangle ,
\end{equation}%
where $x$ and $y$ denote the components of transverse momentum. It should be
noted that here we calculate the two-particle correlator where one particle
is taken from a bin number $1$, and the second particle from a bin number $2$.

Now let us evaluate $\gamma $ at time t=0. In this case it is
reasonable to assume that there is no elliptic anisotropy $v_{2}$ and $%
\gamma =0$ by definition. If we assume that bins are separated enough in
rapidity so that there is no momentum exchange between two bins during the
fireball evolution, i.e., the total transverse momentum $\vec{K}_{1,t}$ is
constant, then $\gamma =0$ also in the final state, even if
subsequently a non-zero $v_{2}$ 
is generated. Of course this simple argument demonstrates only that
having the global TMC we can still obtain a nontrivial dependence of 
$\gamma$ as a function of $\eta _{1}-\eta _{2}$. In Ref. \cite{Pratt:2010zn} 
this problem was studied in detail in the cascade model, where it was shown 
that the satisfactory description of the data can be obtained.

We conclude that although the TMC probably cannot explain the data
completely it gives rise to contributions  which are of the same order 
of magnitude and which exhibit  both the transverse
momentum and rapidity dependence in qualitative and partly quantitative
agreement with the STAR data.

\subsection{AMPT model}

\begin{figure}[h]
\begin{center}
\includegraphics[scale=0.8]{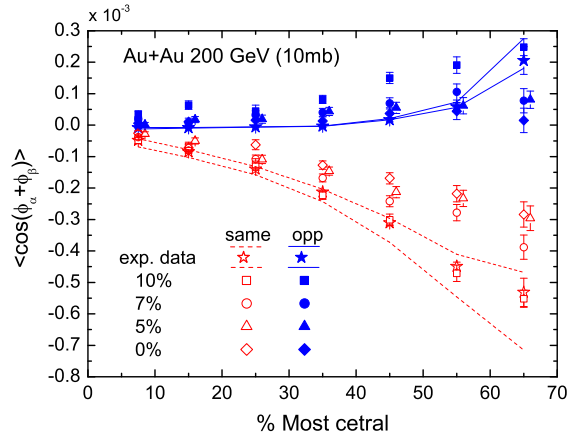}
\end{center}
\caption{The two-particle azimuthal correlator $\protect\gamma $ as a
function of centrality in the AMPT model with different values of initial
charge separation.}
\label{ab_mz_fig1}
\end{figure}

It is desirable  to study the correlators $\gamma $ and $\delta $ also
in an advanced Monte Carlo model, which allows for the 
conservation of transverse momentum on the event-by-event basis, and which generates 
reasonable values for the elliptic
anisotropy $v_{2}$. Recently such calculation was performed in the AMPT
model and the results are presented in Ref. \cite{Ma:2011um} with the
conclusion that the signal coming form AMPT is dominated by TMC. Indeed, the
obtained results, summarized in Figs. \ref{ab_mz_fig1} and \ref%
{ab_mz_fig2}, are in very good agreement with the previous discussion.
As seen in Fig. \ref{ab_mz_fig1}, the calculated correlator $\gamma $ is in
a good qualitative agreement with the data, however it underestimates
the STAR data by a factor of $\sim 2$. It was also shown that initializing the AMPT
calculation with the charge dipole leads to a better description of
the data.

\begin{figure}[h]
\begin{center}
\includegraphics[scale=0.8]{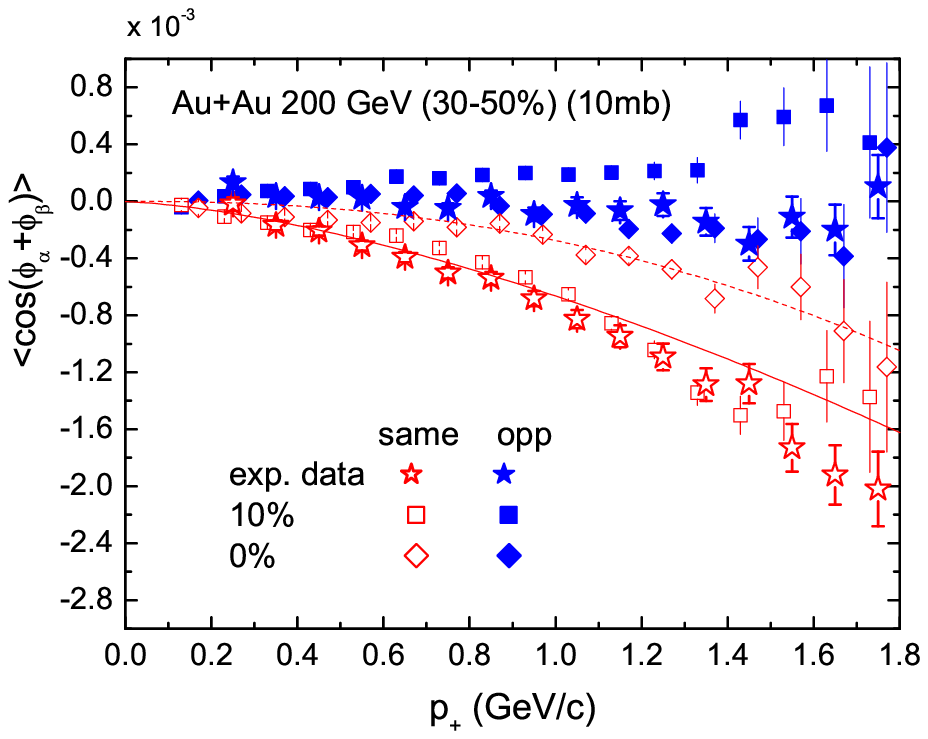} %
\includegraphics[scale=0.8]{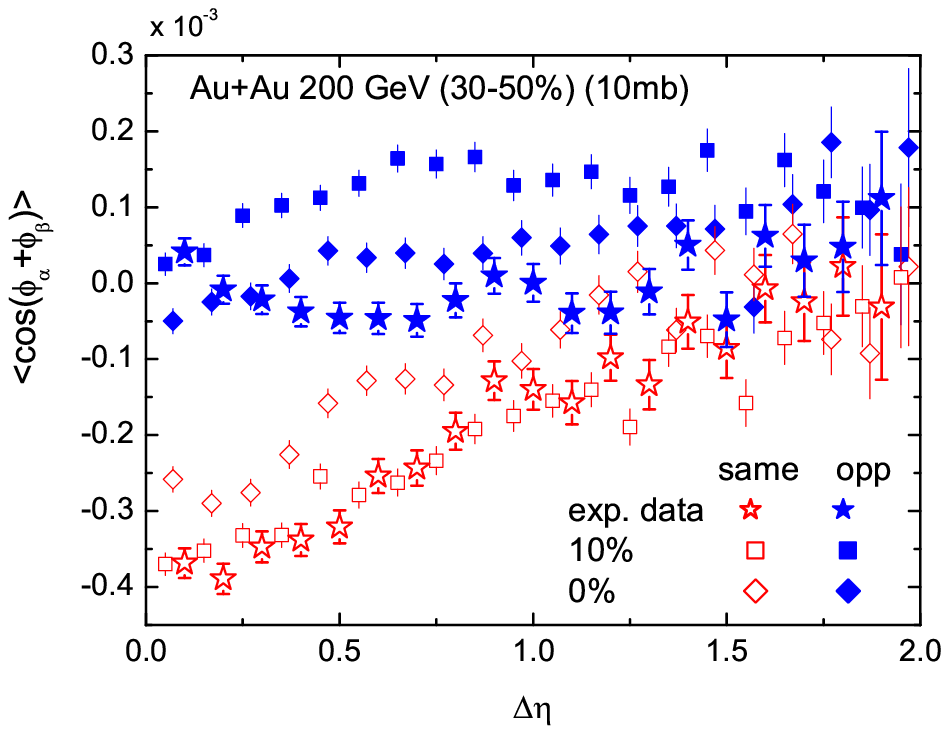}
\end{center}
\caption{The two-particle azimuthal correlator $\protect\gamma $ as a
function of $p_{+}=(p_{1,t}+p_{2,t})/2$ and $\Delta \protect\eta =\protect%
\eta _{1}-\protect\eta _{2}$ in the AMPT model with different initial values
of charge separation.}
\label{ab_mz_fig2}
\end{figure}

In Fig. \ref{ab_mz_fig2} the correlator $\gamma $ is plotted as a function
of $p_{+}=(p_{1,t}+p_{2,t})/2$ and $\Delta \eta =\eta _{1}-\eta _{2}$.
Again, qualitatively the AMPT model reproduce the data but underestimate it
by a factor of $2$. 

As seen from the figures the AMPT model calculation (without initial dipole)
is consistent with the global TMC and allows to understand the behavior of
the STAR data. It clearly demonstrates that all possible background effects
must be studied very thoroughly before any conclusion about local parity
violation can be reached.

\subsection{Local charge conservation}

Given the previous discussion, it is useful to construct a two-particle correlator which is insensitive
to transverse momentum conservation and other charge independent
correlations. The natural choice is the difference of opposite-sign and
same-sign pair correlator (\ref{gam-def}), see \cite{Schlichting:2010qia}:%
\begin{equation}
\gamma _{P}\equiv \frac{1}{2}(2\gamma _{+-}-\gamma _{++}-\gamma _{--}).
\end{equation}%
It is clear that only correlations that are charge sensitive will contribute
to $\gamma _{P}$. While the CME, if present, will contribute to
$\gamma _{P}$,  global TMC, for example will not. 
Therefore the successful description of $\gamma _{P}$ with
conventional physics would constitute a serious challenge for the 
interpretation of the data in terms of the CME.

In Ref. \cite{Schlichting:2010qia} it was argued that $\gamma _{P}$ can be
fully understood assuming that charges are produced later in the collision
(delayed hadronization). Indeed, in the calculation of Ref. \cite{Schlichting:2010qia} 
the charges are produced in pairs in the same point in space-time. 
Due to the collective flow the initial correlation
in space-time is translated into correlations in momentum space, and
consequently it contributes to $\gamma _{P}$. In this approach particles are
emitted according to the blast-wave model with the additional requirement of
local charge conservation at freeze-out. Local charge balance is enforced within the
finite range in rapidity $\sigma _{\eta }$ and the azimuthal angle $\sigma
_{\phi }$. By comparing the model with experimental data on the balance
function \cite{Aggarwal:2010ya}, the values of $\sigma _{\eta }$ and $\sigma
_{\phi }$ can be extracted which allows to make prediction for $\gamma _{P}$.
\begin{figure}[h]
\begin{center}
\includegraphics[scale=0.75]{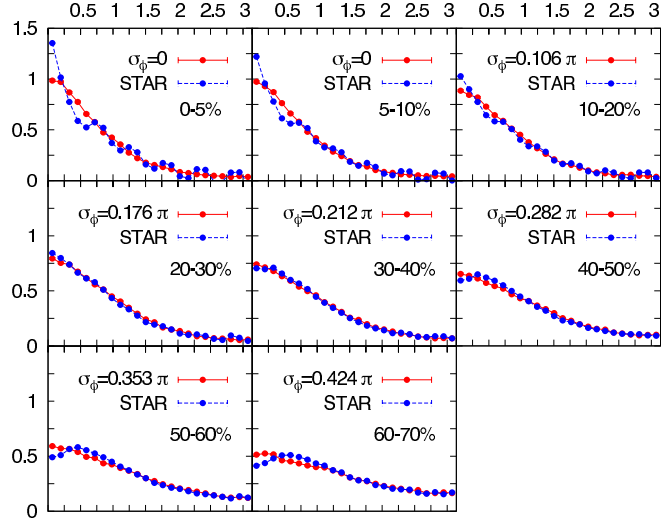}
\end{center}
\caption{Model with a delayed charge production and the local charge
conservation vs the STAR data on the balance function in the relative
azimuthal angle $\Delta \protect\phi =\protect\phi _{1}-\protect\phi _{2}$.}
\label{a_sp_fig3}
\end{figure}

The details of this calculation are presented in Ref. \cite%
{Schlichting:2010qia}. Here we summarize only the main results. It turns out
that the model with delayed charge creation and local charge conservation
can provide a successful description of the balance function both in
the relative angle $\Delta \phi =\phi _{1}-\phi _{2}$ and the relative
pseudorapidity $\Delta \eta $, see Fig. \ref{a_sp_fig3} as an example.%
\footnote{%
Recently a similar model was proposed to explain the fall-off of the
same-side ridge in $\Delta \eta $ \cite{Bozek:2012en}.} 

\begin{figure}[h]
\begin{center}
\includegraphics[scale=0.6]{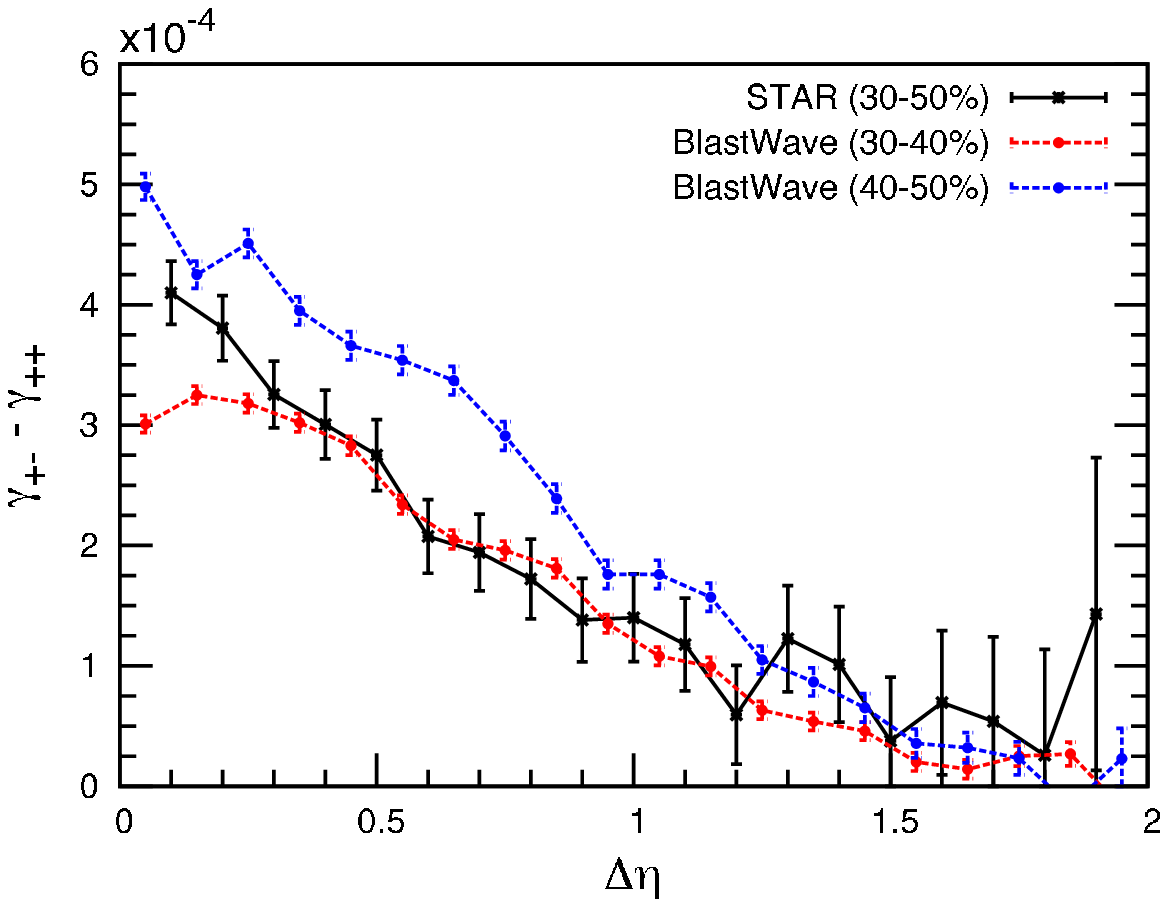} %
\includegraphics[scale=0.6]{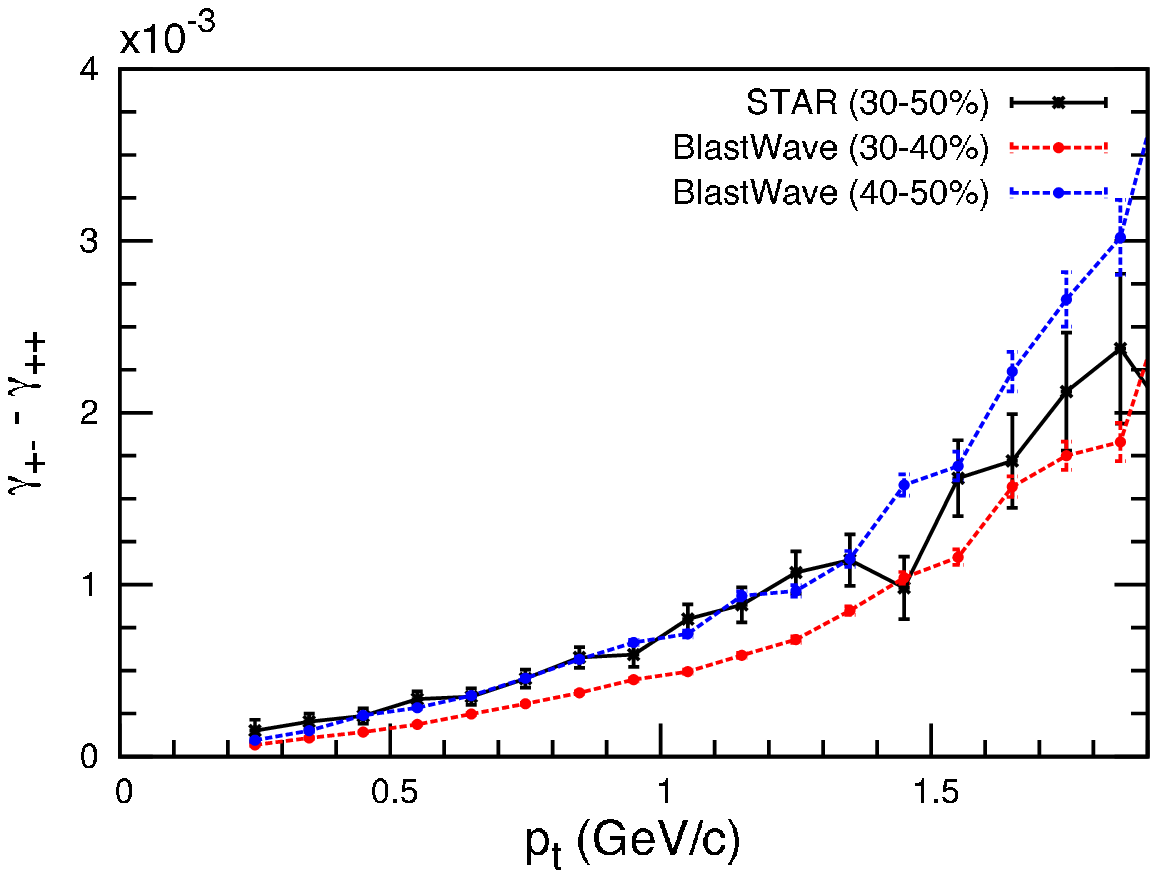}
\end{center}
\caption{Model with a delayed charge production and local charge
conservation vs the STAR data on $\protect\gamma _{P}$ as a function of $%
p_{+}$ and $\Delta \protect\eta $.}
\label{a_sp_fig9}
\end{figure}

Using the best values of parameters $\sigma _{\eta }$ and $\sigma _{\phi }$
the contribution of local charge conservation to $\gamma _{P}$ can be
calculated. The results are presented in Fig. \ref{a_sp_fig9}. 

As seen in Fig. \ref{a_sp_fig9} the agreement of the model with the STAR
data is very good. It suggests that the two-particle charge sensitive
correlations may be dominated by the local charge conservation.

\subsection{Decomposition of flow-induced and flow-independent contributions}

From the analysis of the data in Section 3 as well as the discussion of
various ``background'' effects, it appears rather plausible that the observed
charge-dependent correlation patterns in $\gamma $ and $\delta $ contain
contributions from more than one source. In particular there are
effects whose contributions to these correlations are flow-dependent,
for example 
the transverse momentum conservation (TMC) or the local charge conservation
(LCC). On the other hand the CME, if present, is flow independent.
Let us, therefore, attempt a decomposition of flow-induced and flow-independent
contributions.

We first consider correlation effects where the underlying correlation
function $C$ is independent of the reaction plane orientation: 
\begin{equation}
\mathcal{C}(\phi _{1},\phi _{2})\propto \rho (\phi _{1},\Psi _{RP})\rho
(\phi _{2},\Psi _{RP})C(\phi _{1}-\phi _{2}),
\end{equation}%
where $\rho $ is a single particle distribution. Note that the above is true
for both TMC and LCC effects. 
A correlation effect of this type will
contribute to the measured correlators as follows: 
\begin{equation}
\gamma _{\alpha ,\beta }\sim v_{2}\,F_{\alpha ,\beta }\quad ,\quad \delta
_{\alpha ,\beta }\sim F_{\alpha ,\beta } , \label{gam-F}
\end{equation}%
with the factor $F_{\alpha ,\beta }$ representing the strength of the
effects, and $(\alpha ,\beta )$ is either $++/--$ or $+-$. Both the
TMC and the LCC follow this pattern albeit with opposite
contributions. Thus  $F$ represents the total of all effects of this type.

We should note, however, that the above relation, Eq.~(\ref{gam-F}) is
a simplification of the exact relation between $\gamma $ and
$v_{2}(p_t,\eta)$, which is given in Eq. (\ref{gam-v2}). Since the purpose of
the present discussion is to gain some qualitative insight into the
various contributions, we assume here that $\gamma $ is approximately proportional
to the integrated $v_{2}$.

Next we consider  possible contributions of the CME type. They would
appear  in the two-particle density in the following form
\begin{equation}
\rho _{2}(\phi _{1},\phi _{2})\propto \sin (\phi _{1}-\Psi _{RP})\sin (\phi
_{2}-\Psi _{RP}),
\end{equation}%
which explicitly involves the reaction plane. This term  
will contribute to the measured correlators as follows: 
\begin{equation}
\gamma _{\alpha ,\beta }\sim -H_{\alpha ,\beta }\quad ,\quad \delta _{\alpha
,\beta }\sim H_{\alpha ,\beta },
\end{equation}%
with the factor $H$ representing the strength of the effects. It should be
pointed out that besides the CME, there are possibly other effects that may
also contribute to the correlators with the above pattern. One
example is a possible dipole asymmetry from initial condition fluctuations that
preferably aligns with the out-of-plane direction, see Ref. \cite%
{Teaney:2010vd} for details. However, this effect will be charge
independent, i.e., $H_{++/--}=H_{+-}>0$, whereas the CME predicts a
charge dependence, $H_{++/--}=-H_{+-}>0$. 

\begin{figure}[t]
\begin{center}
\includegraphics[width=5.5cm]{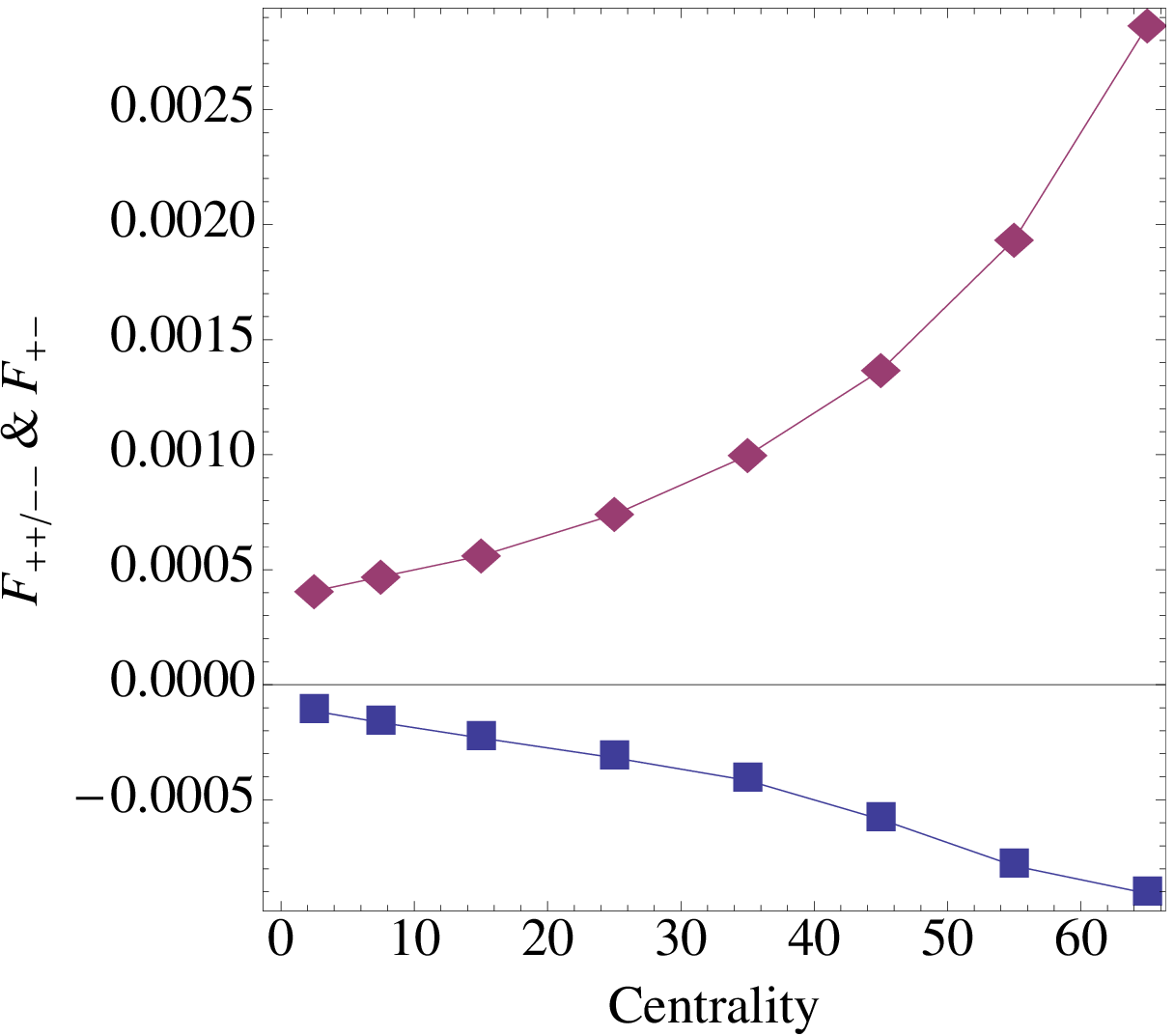}\hspace{0.3cm} %
\includegraphics[width=5.5cm]{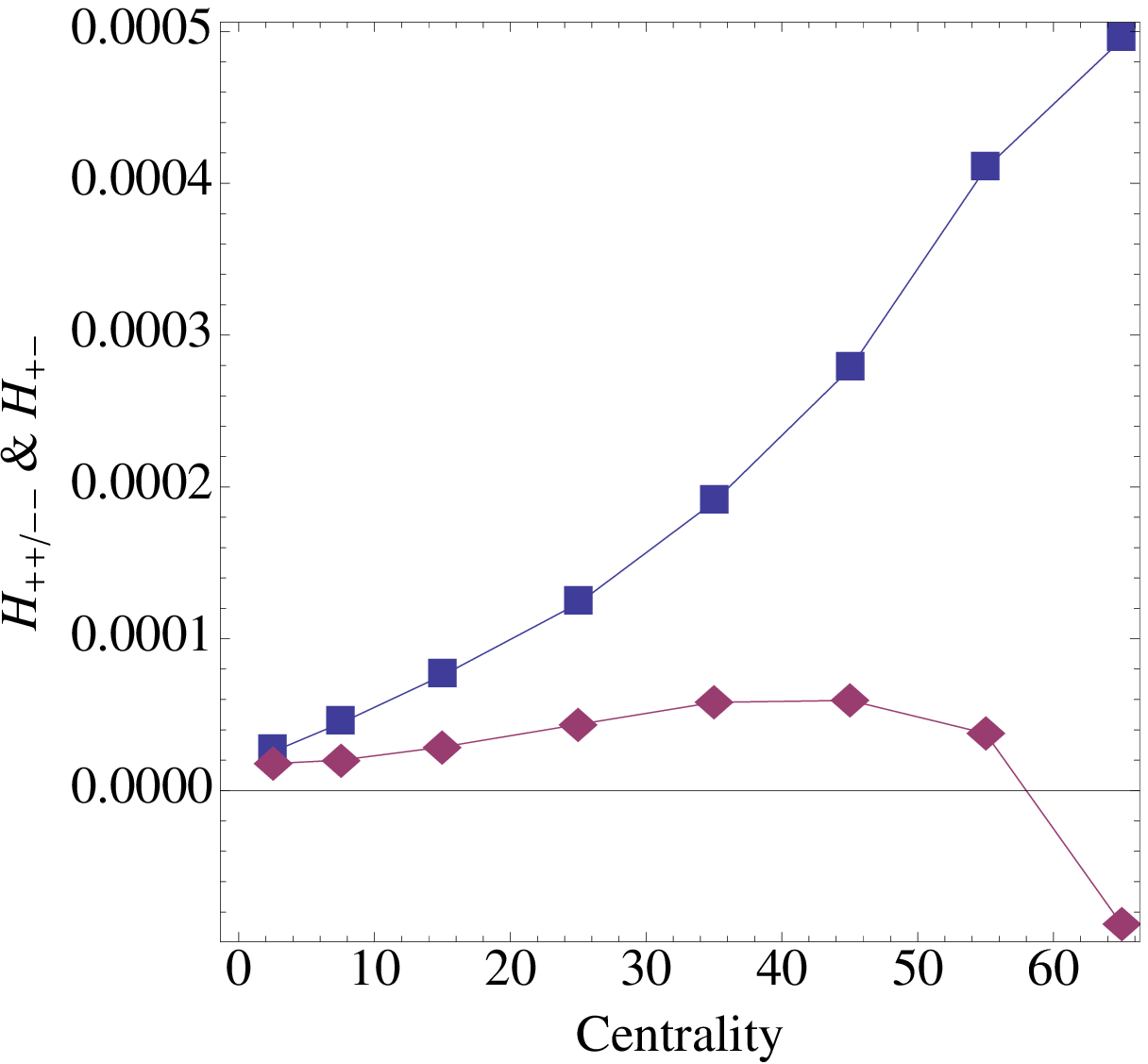}
\end{center}
\caption{The strength factors $F_{\protect\alpha ,\protect\beta }$ (left)
and $H_{\protect\alpha ,\protect\beta }$ (right) extracted from the
decomposition analysis (see text for details). The blue boxes and red
diamonds are for $++/--$ and $+-$, respectively.}
\label{fig_decomposition}
\end{figure}

Combining the two types of contributions we arrive at the following
decomposition for the reaction plane dependent and independent 
correlation functions,
\begin{equation}
\gamma _{\alpha ,\beta }\sim v_{2}\,F_{\alpha ,\beta }-H_{\alpha ,\beta
}\quad ,\quad \delta _{\alpha ,\beta }\sim F_{\alpha ,\beta }+H_{\alpha
,\beta }.
\end{equation}%
Given these relations, we may use the STAR data for $%
\gamma _{\alpha ,\beta }$, $\delta _{\alpha ,\beta }$, and $v_{2}$ to
extract the strength factors $F_{\alpha ,\beta }$ and $H_{\alpha ,\beta }$
as a function of centrality. The result of such a decomposition is shown in
Fig. \ref{fig_decomposition}. 

Given this analysis we make the
following observations: (a) Both components are charge dependent,
i.e. there is significant
difference between $++/--$ and $+-$; (b) In both cases, however, the
same-charge and opposite-charge signals are not symmetric with respect to zero.  This may
indicate that in each category there are likely more than one source of
correlations; (c) There is  a strong residual centrality dependence for
both types component, although the dependence on
centrality from $v_{2}$ has been removed. This may indicate that the
correlations depend also on the multiplicity, which changes from central to
peripheral collisions.

Although, as already noted, the above analysis is qualitative, let us
entertain a possible scenario, which would be consistent with the
above observations:  The flow-induced signals may have
two sources, the TMC with $F_{++/--}^{TMC}=F_{+-}^{TMC}<0$ and the LCC
with $F_{++/--}^{LCC}=0$ and $F_{+-}^{LCC}>|F_{+-}^{TMC}|>0$. The
flow-independent signals may be from  two different sources, the CME with $%
H_{++/--}^{CME}>0$ and $H_{+-}^{CME}<0$ and the dipole asymmetry from
fluctuations (DAF) with $H_{++/--}^{DAF}=H_{+-}^{DAF}>0$. Such a
combination would indeed lead to correlations with magnitude and
sign in qualitative agreement with the data. However, a quantitative
analysis would have to be based on the exact decomposition based on 
Eq.~(\ref{gam-v2}). Alternatively, one may attempt a separation
of flow dependent and flow independent contributions in
experiment. How this could be achieved will be discussed in the following.

\subsection{Suppression of elliptic-flow-induced correlations}

In this chapter we will discuss the possibility of removing the
elliptic-flow-induced background from the experimental data.

As seen in Eq. (\ref{gam-v2}) all contributions due to correlations are proportional to
the elliptic flow parameter, $v_2$. Therefore, it would be desirable
to control or remove this contribution by a suitable
measurement. There are essentially two ways to go about this. 

First, as
proposed in Ref. \cite{Voloshin:2010ut}, is to study 
collisions of deformed nuclei, such as $U+U$. By selecting very
central, ``face on face" collisions where the deformation of the
nuclei is imprinted on the fireball, elliptic flow should be generated
while at the same time the magnetic field will be very small. 
Should one observe correlations of the same magnitude and
structure as ones already reported by STAR, this would identify their
origin as being due to conventional two-particle correlations. The
observation of considerably smaller correlations combined with a
sizable $v_2$, on the other hand, would lend support
for the existence of the CME. This approach, while challenging to analyze,
is at present being attempted at RHIC, were first $U+U$ collisions were
made available.

\begin{figure}[h]
\begin{center}
\includegraphics[scale=0.4]{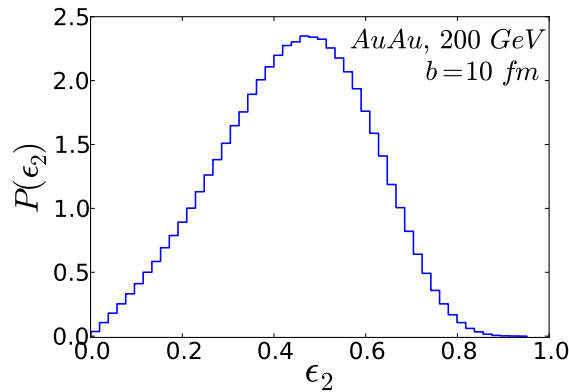}
\end{center}
\caption{The distribution of initial eccentricity $\protect\epsilon _{2}$
calculated in the Glauber Monte-Carlo at the impact parameter $b=10$ fm.}
\label{ab_e21}
\end{figure}

Alternatively, as proposed  in Ref. \cite{Bzdak:2011my} one may make use of 
% the well established
the large event-by-event fluctuations of $v_2$. By selecting events with different $v_{2}$ in a given
centrality class we can control this background. In principle
the measurement can be extrapolated to $v_{2}=0$ (and consequently 
$v_{2}(p_{t},\eta )=0$) which will allow to extract correlations that only
depend on the reaction plane orientation. Indeed, as presented in 
Fig. \ref{ab_e21} even at $b=10$ fm we expect a large fluctuation of initial
eccentricity that will translate to large fluctuations of elliptic flow $v_{2}$. 

\begin{figure}[h]
\begin{center}
\includegraphics[scale=0.4]{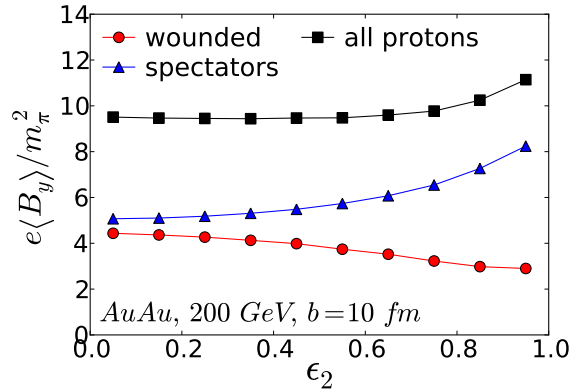}
\end{center}
\caption{The out-of-plane component of the magnetic field from wounded
protons and spectator protons as a function of initial eccentricity $\protect%
\epsilon _{2}$ at a given impact parameter $b=10$ fm. }
\label{ab_By1}
\end{figure}

Of course it is important to remove this background under the condition that
the contribution from the Chiral Magnetic Effect is approximately unchanged.
In Fig. \ref{ab_By1} we demonstrate that indeed it is the case. Both the
wounded nucleons and spectators' contribution to the magnetic field weekly
depends on $\epsilon _{2}$. 

We believe this analysis should help to clarify the situation. Observation
of non-zero $\gamma _{++}$ or $\gamma _{+-}$ at vanishing value of elliptic
anisotropy $v_{2}$ will suggest the existence of the correlation mechanism
that is sensitive to the average direction of the 
magnetic field -- possibly the Chiral Magnetic Effect.

To summarize this Section, clearly there are many contributions based
on conventional physics which contribute to the azimuthal correlations
analyzed by the various experiments. In addition to those discussed in
more detail in this Section other mechanisms, such as the decay of
multi-particle clusters \cite{Wang:2009kd}, have been proposed. While
it is more difficult to asses their contribution quantitatively, their
influence cannot a priory be ignored. Therefore, it seems the best way
forward is to separate the influence of elliptic flow and magnetic
field experimentally as discussed in the last part of this Section. 

\section{Summary and Conclusions}
\label{sec:5}

In this review we have concentrated on the observational aspects of the
search for phenomena related to local parity violation of the strong interaction. 
Specifically we have discussed various observables and their measurement for the
charge separation, which is a predicted consequence of induced currents
due to sphaleron and anti-sphaleron transitions in an external magnetic field. 
This phenomenon is
often referred to as the Chiral Magnetic Effect (CME). 

We have discussed various properties and aspects of azimuthal angle
correlations, and we have emphasized that, due to the elliptic flow
observed in heavy ion collisions, virtually any two-particle
correlations contribute to the azimuthal correlations. We have further
discussed an alternative observable, which in our view may be better
suited in discriminating between the backgrounds and the CME. 

We have examined the presently available data on reaction
plane dependent and independent correlation functions of same- and
opposite-charged pairs. Our phenomenological analysis of the data by
the STAR collaboration showed that the measured correlations of same-charge 
pairs are predominantly back-to-back, and in-plane. This is opposite to the
predictions from the CME, where same-side out-of-plane correlations
for pairs of the same charge are expected. The data by the ALICE
collaboration taken at about ten times the STAR collision energy, on the other hand, show a correlation, albeit
small, which is qualitatively consistent with the CME expectations. 

However, before
any conclusion on the CME can be drawn, the contributions due to
``conventional'' correlations need to be accounted for. As we have
discussed in some detail, both the conservation of transverse momentum
as well as local charge conservation give rise to corrections which
are of the same order as the experimental signal. These need to be
understood and properly subtracted from the data in order to see if a
signal consistent with the CME remains. 
Since there are conceivably many
other two-particle correlations, which may enter due to the presence
of elliptic flow, an important step towards answering the question about
the existence of the CME is to experimentally 
disentangle the elliptic flow phenomenon
from the creation of a strong magnetic field. This can be either done by
colliding deformed nuclei or by carefully utilizing the fluctuations
of the elliptic flow.

In conclusion, the present experimental evidence for the existence of
the CME is rather ambiguous. While progress on the assessment of the
various background terms is to be expected, the sheer variety of
possible correlations will likely limit a reliable quantitative
determination of all the backgrounds. Therefore, the most important
next step is the experimental separation of elliptic flow and
magnetic field.
In this context it is encouraging to note, that the
first $U+U$ collisions at RHIC have just been recorded.

Finally let us close with a note of caution. Besides the CME there
are other phenomena related to local non-vanishing topological
charge fluctuations, such as the Chiral Magnetic Wave. One of the
predictions in this case is the difference of elliptic flow between
positively and negatively charged pions for collisions at lower
energies \cite{Burnier:2011bf}. However, again there may be other,
more mundane effects which lead to similar phenomena, such as an
increased stopping of baryon number and isospin at lower energies
\cite{Steinheimer:2012bn}.

\bigskip

\textbf{Acknowledgments}

A.~B. was supported by Contract No. DE-AC02-98CH10886 with the U. S. Department of
Energy. V.~K. was supported by the Office of Nuclear Physics in the US
Department of Energy's Office of Science under Contract No.
DE-AC02-05CH11231. J.~L. is grateful to the RIKEN BNL Research Center for partial support. A.~B. also acknowledges the grant No. N202 125437 of the Polish Ministry of Science and Higher Education (2009-2012).

\end{document}